\documentclass[pdftex, epjc3]{svjour3} 
\RequirePackage[T1]{fontenc}
\RequirePackage{widetext}
\RequirePackage{graphicx}
\RequirePackage{flushend}
\RequirePackage[numbers, sort&compress]{natbib}
\RequirePackage[colorlinks, citecolor=blue, urlcolor=blue, linkcolor=blue]{hyperref}
\RequirePackage{amsmath}
\RequirePackage{amssymb}
\RequirePackage{latexsym}
\RequirePackage{color}
\RequirePackage{bm}
\RequirePackage{multirow}
\RequirePackage{cancel}
\RequirePackage{hyperref}
\RequirePackage{inputenc}
\journalname{Eur. Phys. J. C}
\begin{document}
\title{Mono-Higgs Signature in the Scotogenic Model with Majorana Dark Matter}

\author{Amine~Ahriche \thanksref{e1, ad1, ad2}, Abdesslam~Arhrib \thanksref{e2, ad3}, Adil~Jueid \thanksref{e3, ad4}, Salah~Nasri \thanksref{e4, ad5, ad2} \& Alejandro~de~la~Puente \thanksref{e5, ad6} }

\thankstext{e1}{aahriche@ictp.it}
\thankstext{e2}{aarhrib@gmail.com}
\thankstext{e3}{adil.jueid@sjtu.edu.cn}
\thankstext{e4}{snasri@uaeu.ac.ae}
\thankstext{e5}{lagrange2001@gmail.com}

\institute{Department of Physics, University of Jijel, PB 98 Ouled Aissa, DZ-18000
Jijel, Algeria. \label{ad1}
\and
The Abdus Salam International Centre for Theoretical Physics, Strada
Costiera 11, I-34014, Trieste, Italy.\label{ad2}
\and
D\'{e}partement de Math\'{e}matiques,
Facult\'{e} des Sciences et Techniques,
Universit\'{e} Abdelmalek Essaadi, B. 416, Tangier, Morocco. \label{ad3}
\and
INPAC, Shanghai Key Laboratory for Particle Physics and Cosmology,
Department of Physics and Astronomy, Shanghai Jiao Tong University,
Shanghai 200240, China. \label{ad4}
\and
Department of physics, United Arab Emirates University, Al-Ain, UAE. \label{ad5}
\and
New York Academy of Science, 7 World Trade Center, 250 Greenwich St. 40th Floor
New York, NY 10007-2157, USA. \label{ad6}}

\maketitle

\begin{abstract}
We study the phenomenology of scotogenic model in the case of Majorana Dark Matter (DM) candidate. This scenario gives
important consequences since the parameter space of the model is almost unconstrained compared to the Inert Higgs Doublet Model
(or the scotogenic model with scalar DM), and hence, offers new opportunities for discovery at future high energy collider,
e.g. the HL-LHC. As an example, we focus on the production of the Standard Model (SM) Higgs boson
in association with a pair of dark scalars. Owing to its clean signature,
the $\gamma\gamma$ decay channel of the SM Higgs boson is investigated
in great detail at both the HL-LHC (at $\sqrt{s}=14$ TeV) and the future
FCC-hh (at $\sqrt{s}=100$ TeV). After revisiting the LHC constraints from
run-II on the parameter space of the model, and selecting benchmark points satisfying
all the theoretical and experimental constraints, we found that scalars with mass up to
$140$ GeV ($160$ GeV) can be probed at the LHC (FCC-hh) with a
$3$ ab$^{-1}$ of integrated luminosity assuming $5\%$ of uncertainty.
\end{abstract}

\section{Introduction}
\label{sec:introduction}

The observation of neutrino oscillations in solar, atmospheric, reactor and accelerator experiments remains one clear
indication that the Standard Model (SM) is not a complete framework of fundamental physics. The smallness of the observed
neutrino masses tells that at the non-renormalizable level we might not have a straightforward answer to the mechanism that
bestows neutrinos with mass. One popular mechanism for generating tiny neutrino mass is the so called seesaw
mechanism~\cite{Mohapatra:1979ia, Schechter:1980gr, Schechter:1981cv}. However, realistic models based on the seesaw
mechanism involve high mass scales that are hard to be probed at collider experiments. Neutrino mass generation through
loop diagrams is interesting and give \emph{naturally} small masses due to loop-suppression factors. Therefore, these
models can be probed at present and future colliders. In these class of models, the smallness of neutrino mass has been
addressed within frameworks at one-loop~\cite{Zee:1985rj, Ma:1998dn},
two loops~\cite{Zee:1985id, Babu:1988ki, Aoki:2010ib, Guo:2012ne, Kajiyama:2013zla},
three loops~\cite{Krauss:2002px, Aoki:2008av, Aoki:2009vf, Gustafsson:2012vj, Ahriche:2014cda, Ahriche:2014oda, Hatanaka:2014tba, Nishiwaki:2015iqa,
Ahriche:2015wha, Ahriche:2015loa, Okada:2015hia, Nomura:2016ezz, Gu:2016xno, Cheung:2016frv, Cheung:2017efc, Dutta:2018qei}, and four loops~\cite{Nomura:2016fzs}.

Additionally, experimental evidence of dark matter (DM) has driven many years of investigation shedding light on possible particle and
electroweak-size interaction explanations that can reproduce the observed DM relic abundance in the Universe. This paradigm is interesting
since it can be tested at colliders such as the Large Hadron Collider (LHC). One of the simplest extensions of the SM consists in incorporating
an additional Inert Higgs Doublet $\Phi$ with a discrete $Z_{2}$ symmetry under which the new scalar is odd, $\Phi\to-\Phi$, and the other SM
fields even~\cite{Deshpande:1977rw}. In this case, the lightest odd particle would act as DM candidate.
This model, known as the Inert Higgs Doublet Model (IHDM), contains one CP-even Higgs identified as the SM Higgs, an other
CP-even Higgs $H^0$, one CP-odd $A^0$ and a pair of charged Higgs $H^\pm$, and consequently has a rich
phenomenology~\cite{Barbieri:2006dq, Ma:2006km, Gustafsson:2007pc, Hambye:2007vf, Agrawal:2008xz, Dolle:2009fn, Andreas:2009hj, Dolle:2009ft, Miao:2010rg, Arhrib:2012ia, Arhrib:2013ela, Goudelis:2013uca, Arhrib:2014pva, Arhrib:2015hoa, Castilla-Valdez:2015sng, Kanemura:2016sos, Banerjee:2016vrp, Poulose:2016lvz, Huang:2017rzf, Wan:2018eaz}.
For example, the model provides mono-jet, mono-Higgs, mono-Z, mono-photon signatures that can be tested at the LHC and
future colliders. It appears from the above phenomenological studies that the IHDM is strongly constrained from direct
and indirect DM searches both for low and intermediate DM masses~\cite{Arhrib:2013ela, Belyaev:2018ext}. For DM lighter than 62.5 GeV,
LHC data also puts severe constraints on the invisible decay of the SM Higgs which in turn translate into constraints on a
combination of the scalar parameters of the potential~\cite{Arhrib:2013ela, Belyaev:2016lok}.
Moreover, collider bounds on the IHDM are obtained as a reinterpretation of neutralinos and charginos pair production
both from LEP II~\cite{Lundstrom:2008ai} and from LHC~\cite{Belanger:2015kga}. From LEP II data, Ref.~\cite{Lundstrom:2008ai}
sets an upper bound on the pseudo-scalar mass, $m_{A^{0}}$ (resp $m_{H^{0}}$) , below $100$ GeV (resp $80$ GeV) consistent
with mass splittings $\Delta m(A^{0}, H^{0})\ge 8$ GeV. While from LHC data, Ref~\cite{Belanger:2015kga} limits have been derived
using a dilepton plus missing energy signature which excludes masses for the exotic scalar up to $62.5$ GeV.
A recent study~\cite{Belyaev:2016lok} showed that the LHC at $13$ TeV and $3000$ fb$^{-1}$ luminosity could exclude exotic
scalar masses below $83$ GeV using the mono-jet channel.

However, If one focuses on a degenerate spectrum of exotic scalars, which is a natural outcome of accidental symmetries
in the scalar potential~\cite{Blinov:2015qva}, the region of scalar masses above $M_Z/2$ remains unconstrained for splittings
between the exotic scalar and the charged scalar mass below $5$ GeV. It was also found that LHC searches are not strong enough
to probe the degenerate window due to lepton $p_{T}$ requirements. In the light of current collider experimental bounds and the
viable region of parameter space in the IHDM, and in order to address the DM nature, one has to go beyond this minimal extension
of the SM. For instance, extending the IHDM by three right handed Majorana fermions may provide a possible solution to the problem
of over-constrained quartic couplings and, on the other hand, give rise to small neutrino masses generated through one-loop diagrams.
 In the present work, we build on a recent phenomenological analysis in the framework of scotogenic model~\cite{Ma:2006km} performed
by some of us~\cite{Ahriche:2017iar}\footnote{The phenomenology of the scotogenic model has been extensively studied in the
literature~\cite{Merle:2015gea, Kitabayashi:2018bye, Hugle:2018qbw, Baumholzer:2018sfb, Borah:2018rca}.}. The scotogenic model is a SM
minimal extension where the SM neutrinos obtain naturally small masses at the one-loop order. In order to achieve this, the scalar potential
has to be augmented by an inert complex scalar doublet with a small mixing quartic coupling to the SM Higgs. Due to the new
Yukawa couplings, the scalar potential has an enhanced $SU(2)$ symmetry acting only on the exotic scalar and the new right-handed neutrino fields
\footnote{This symmetry, however, is broken explicitly by the Majorana bare mass terms}. Because of this global symmetry, the quartic coupling $\lambda_5$
between $\Phi$ and the SM Higgs, that is responsible for the mass splitting between the CP-odd and CP-even neutral scalars, does not run and
thus can naturally be very close to zero, which naturally yields small mass for the active neutrino. In contrast to the
 $\lambda_5$ term in the potential, there is another coupling between $\Phi$ and the SM Higgs, $\lambda_4$, which has non-vanishing $\beta$-function even if the coupling is chosen to be zero at some very high energy scale\footnote{Here, one notices that the global $SU(2)$ is radiatively broken to a global $U(1)$ that leads to the degeneracy between the CP-odd and CP-even scalars.}. This region of parameter space corresponds also to a spectrum of a compressed exotic scalar/pseudo-scalar spectrum that leads to interesting collider signatures which are difficult
to probe in the IHDM with current and near-future data.

In~\cite{Ahriche:2017iar, Vicente:2014wga}, the DM candidate is considered to be the lightest Majorana fermion $N_1$, which
implies significant difference in the parameter space compared to both IDHM or the scotogenic model with scalar DM candidate.
For instance, in case of scalar DM candidate, the null results from searches in direct detection imply that the coupling combination
$\lambda_L=\lambda_3+\lambda_4+\lambda_5$ is extremely suppressed
to be suppressed, while for the fermionic DM case this constraint do not affect the scalar potential's parameters. In the fermionic DM case,
the CP-odd and CP-even scalars decay predominantly into SM neutrino and the Majorana fermion $N_i$, and therefore they can not be seen at colliders, i.e.,
they behave as dark scalars. In other words, both the IHDM and the scotogenic model provide identical signatures at colliders but
with different event yields since they have different parameter space.

As pointed above, the production of dark scalars can lead to several signatures dubbed as mono-X. The most known of and studied in the literature is the
mono-jet signature. However, within the framework of scotogenic model, the mono-jet signature is only sensitive to the masses of the particles produced
in the final state and not to the scalar couplings such as e.g. $\lambda_L$. The reason for this is that the mono-jet cross section gets the most important
contribution from diagrams with the exchange of $Z$-boson and involving \emph{gauge couplings} only. Therefore, alternatives to the mono-jet channel need
to be exploited. In this regard, we focus the scope of this work on the mono-Higgs channel in the diphoton final state at both the HL option of the LHC
at $14$ TeV and a Future Circular Collider (FCC-hh) at $100$ TeV. This signature is an excellent probe of new physics and
DM~\cite{Carpenter:2013xra, Petrov:2013nia, Abdallah:2016vcn, Ghorbani:2016edw}. We stress out that searches of DM in events with Higgs
and missing transverse energy have been carried out by the ATLAS and the CMS
collaborations~\cite{Aad:2015yga, Aaboud:2017uak, Aaboud:2016obm, Aaboud:2017yqz, Sirunyan:2017hnk, Sirunyan:2018gdw, Sirunyan:2018fpy}
using $\tau^+ \tau^-, \gamma \gamma$ and $b\bar{b}$ decay channels of the Higgs boson. These searches yielded null results which were used
to put severe constraints on simplified models of DM production at hadron colliders. However, these limits do not apply to our model due to
the smallness of the corresponding production cross sections of DM particles in association with a Higgs boson. In this work, we will follow
closely the analysis setup of reference~\cite{Aaboud:2017uak}. \\
The outline of the paper is as follow. In the second section we review the scotogenic model and all the theoretical and the
experimental bounds in the entire degenerate window, where all exotic scalars have approximate equal masses. We then carry out a
complete comparison of this model to the latest LHC run II data, and expose the available parameter space in the third section.
In the fourth and the fifth sections, we present a full sensitivity analysis to a mono-Higgs signature within this framework.

\section{The Model: Parameters and Constraints}
\label{sec:model}
\subsection{Model}
In this model, the SM is extended by one $SU(2)_L$ inert Higgs doublet and three singlet Majorana fermions
$N_{i}\sim(1, 1, 0)$, $i=1, 2, 3$. These new particles are odd under a $Z_2$ symmetry, whereas the SM particles are even. In this setup, the most general gauge-invariant, and renormalizable scalar potential that is invariant under CP- and $Z_2$-symmetries has the form
\begin{eqnarray}
V & = & -\mu_{1}^{2}|H|^{2}+\mu_{2}^{2}|\Phi|^{2}+\frac{\lambda_{1}}{6}|H|^{4}+\frac{\lambda_{2}}{6}|\Phi|^{4} +\lambda_{3}|H|^{2}|\Phi|^{2} + \frac{\lambda_{4}}{2}|H^{\dagger}\Phi|^{2} + \frac{\lambda_{5}}{4}\left[(H^{\dagger}\Phi)^{2}+h.c.\right], \label{V}
\end{eqnarray}

The electroweak symmetry breaking occurs due to the non vanishing Vacuum Expectation Value (VEV)
acquired by the SM Higgs doublet, through its neutral component, while the $Z_2$-odd inert doublet $\Phi$
does not develop a VEV as its quadratic term has positive curvature. The SM Higgs and the inert doublets can be parametrized as
\begin{equation}
H=\left(\begin{array}{c}
G^{+}\\
\frac{1}{\sqrt{2}}(\upsilon+h+iG^{0})
\end{array}\right), \, \Phi=\left(\begin{array}{c}
H^{+}\\
\frac{1}{\sqrt{2}}(H^{0}+iA^{0})
\end{array}\right).
\end{equation}

The Lagrangian that involves the Majorana fermions can be written as
\begin{eqnarray}
\mathcal{L} & \supset & h_{ij}\bar{L}_{i}\epsilon\Phi N_{j}+\frac{1}{2}M_{i}\bar{N}_{i}^{C}N_{i}+h.c., \label{LL}
\end{eqnarray}
where $\bar{L}_{i}$ is the left-handed lepton doublet and $\epsilon=i\sigma_{2}$
is an antisymmetric tensor. Note that the absence of $\bar{L}_{i}HN_{j}$ in
the Lagrangian (\ref{LL}) is due to the imposed discrete $Z_{2}$
symmetry. The parameters $\lambda_{1}$ and $\mu_{1}^{2}$
in (\ref{V}) can be eliminated in favor of the SM Higgs mass and
its VEV ($\upsilon=246$ GeV), which is considered
at one-loop level \`a la $\overline{\mathrm{DR}}$ scheme~\cite{Martin:2001vx}. After EWSB, three degrees of freedom are absorbed by the longitudinal
gauge bosons and we are left with two CP-even scalars $(h^0, H^{0})$, one CP-odd scalar
$A^{0}$ and a pair of charged scalars $H^{\pm}$. Their tree-level masses are given by:
\begin{eqnarray}
m_{H^{\pm}}^{2} = \mu_{2}^{2}+\frac{1}{2}\lambda_{3}\upsilon^{2}, ~m_{H^{0}, A^{0}}^{2} = m_{H^{\pm}}^{2}+\frac{1}{4}\left(\lambda_{4}\pm\lambda_{5}\right)\upsilon^{2}.
\end{eqnarray}

\begin{figure}[h]
\begin{center}
\includegraphics[width=0.48\textwidth]{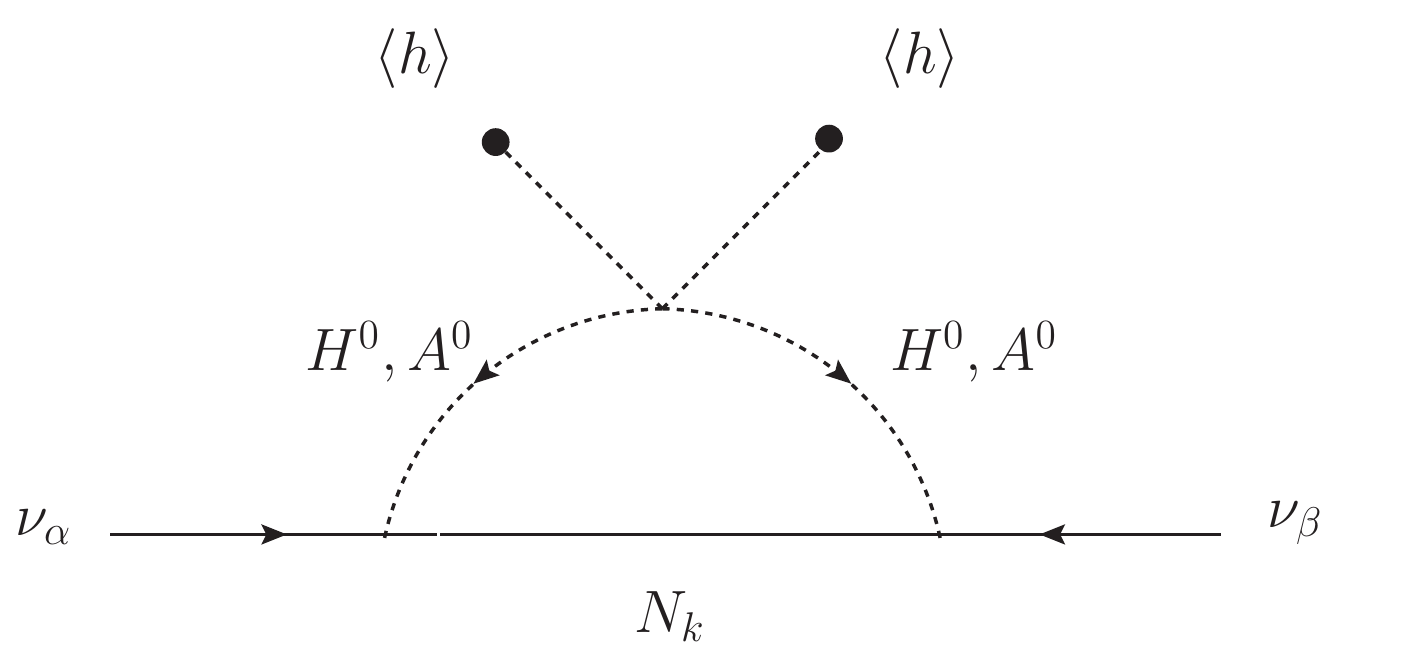}
\caption{Feynman diagram responsible for the neutrino mass.}
\end{center}
\label{fig:Nu}
\end{figure}

The neutrino mass can be obtained at the one loop level via the diagram in Fig.~\ref{fig:Nu}. The neutrino mass matrix
elements~\cite{Ma:2006km, Cai:2017jrq} are given \`a la Casas-Ibarra form~\cite{Casas:2001sr} by
\begin{align}
m_{\alpha\beta}^{(\nu)} = & \sum_{k}h_{\alpha k}.G_k.h_{k\beta}^{T}.\nonumber\\
G_k = & \frac{M_{k}}{16\pi^{2}}\left\{ \frac{m_{H^{0}}^{2}}{m_{H^{0}}^{2}-M_{k}^{2}}\ln\frac{m_{H^{0}}^{2}}{M_{k}^{2}}-\frac{m_{A^{0}}^{2}}{m_{A^{0}}^{2}-M_{k}^{2}}\ln\frac{m_{A^{0}}^{2}}{M_{k}^{2}}\right\}\label{Mnu}
\end{align}

In this model, the smallness of neutrino mass is a consequence
of the tiny mass splitting in the inert neutral sector. In other words, the following ratio
$\epsilon: = \frac{|\lambda_{5}|\upsilon^{2}}{m_{H^{0}}^{2}+m_{A^{0}}^{2}}$ is much smaller than unity.
Then, after the expansion over $\epsilon$, the parameter $G_k$ in (\ref{Mnu}) is given by

\begin{eqnarray}
G_k = \frac{|\lambda_{5}|}{16\pi^{2}}\frac{\upsilon^{2}}{\bar{m}}\left[\frac{x_{k}}{1-x_{k}^{2}}+\frac{x_{k}^{3}}{\left(1-x_{k}^{2}\right)^{2}}\ln x_{k}^{2}\right],
\end{eqnarray}

with $x_{k}=M_{k}/\bar{m}$ and $\bar{m}^{2}=(m_{H^{0}}^{2}+m_{A^{0}}^{2})/2$.
According to the Casas-Ibarra parameterization, the coupling $h$ can be written as
\begin{equation}
h=D_{\sqrt{G_k}}\mathcal{R}D_{\sqrt{m_{\nu}}}U_{\nu}^{\dag},
\end{equation}
where $D_{\sqrt{G_k}}=\mathrm{diag}\left\{ \sqrt{G_1}, \sqrt{G_2}, \sqrt{G_3}\right\}$, and $D_{\sqrt{m_{\nu}}}=\mathrm{diag}\left\{ \sqrt{m_{1}}, \sqrt{m_{2}}, \sqrt{m_{3}}\right\}$, $\mathcal{R}$\ is an orthogonal rotation matrix ($m_{1, 2, 3}$ are
the neutrino eigenmasses), and $U_{\nu}$ is the Pontecorvo-Maki-Nakawaga-Sakata
(PMNS) mixing matrix~\cite{Pontecorvo:1967fh}. The parameters of the model are subject to constraints from the measurements of the mixing angles and mass-squared differences~\cite{Tortola:2012te} which we implement in our analysis.

\subsection{Constraints}

The parameters of the scalar potential have to satisfy a number of theoretical
and experimental constraints. On the theoretical side, we should require
 perturbativity of all the quartic couplings of the scalar fields.
In addition, the scalar potential has to be bounded from below
in all directions of the field space. For that, the necessary and sufficient conditions are given by~\cite{Branco:2011iw}

\begin{eqnarray}
\lambda_{1, 2} > 0, ~\lambda_3 + \lambda_4 -|\lambda_5| + 2\sqrt{\lambda_1 \lambda_2} >0~, \lambda_3+2\sqrt{\lambda_1 \lambda_2} > 0 \; .
\label{eq:lamCONS}
\end{eqnarray}

However, these constraints do not ensure the vacuum stability since the inert vacuum may not be the global minimum of the potential, and to guarantee
this feature we should also impose the condition $\frac{\mu_{1}^{2}}{\sqrt{\lambda_{1}}}\geq-\frac{\mu_{2}^{2}}{\sqrt{\lambda_{2}}}$~\cite{Ginzburg:2010wa}. Another set of constraints comes from the tree-level perturbative unitarity which which should be preserved at high energies in variety of processes
involving scalars or gauge bosons. At high energies, using the equivalence theorem, we replace the longitudinal $W$ and $Z$ bosons by the corresponding
charged and neutral Goldstone bosons respectively. Therefore, we are left only with pure scalar scattering amplitudes. Computing the decay amplitudes for these processes, one finds
a set of $4$ matrices with quartic couplings as their entries. The
eigenvalues for those matrices have to be smaller than $4\pi$~\cite{Kanemura:1993hm, Akeroyd:2000wc}.

\begin{figure}[!h]
\includegraphics[width=0.48\textwidth]{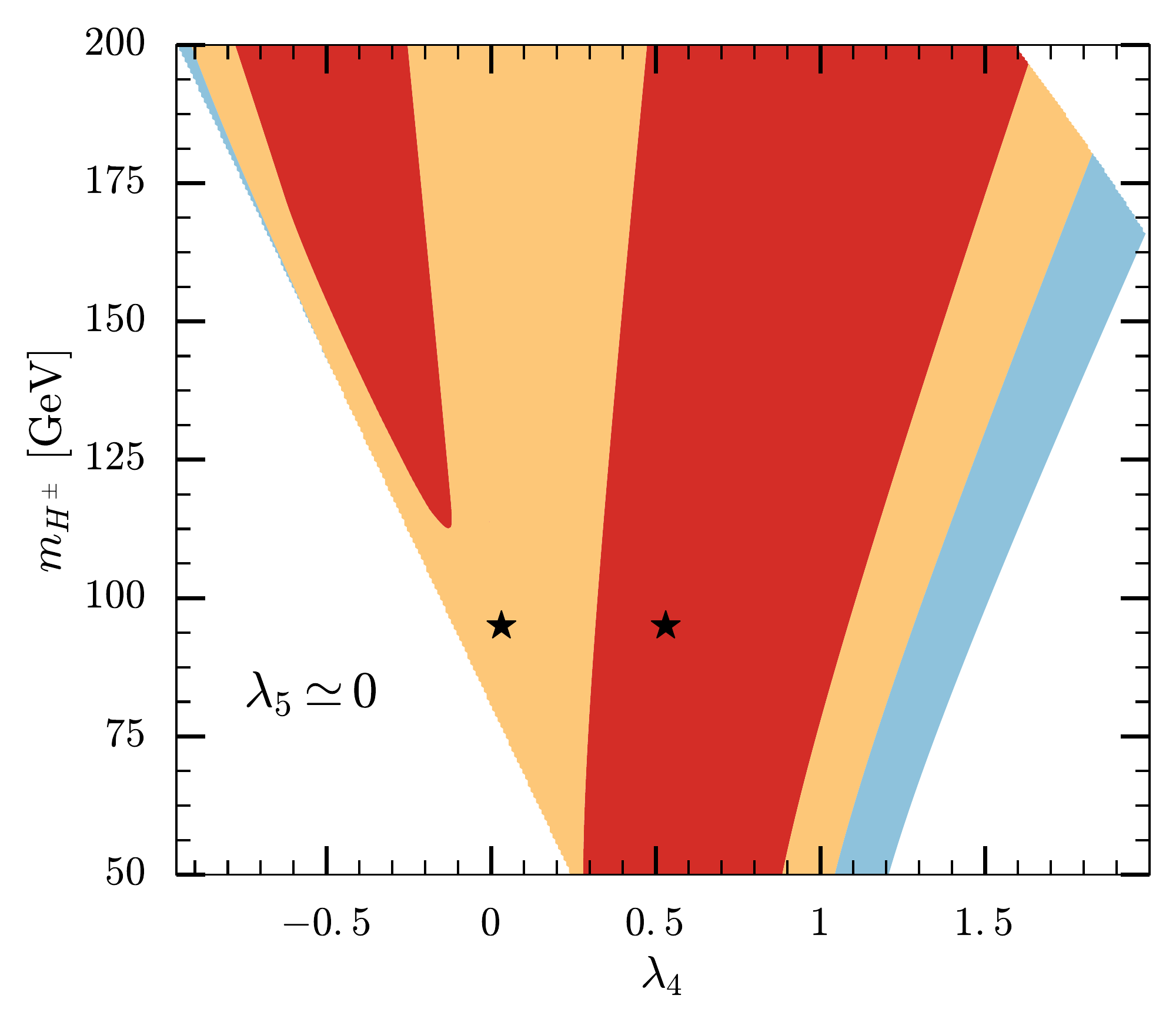}
\hfill
\includegraphics[width=0.48\textwidth]{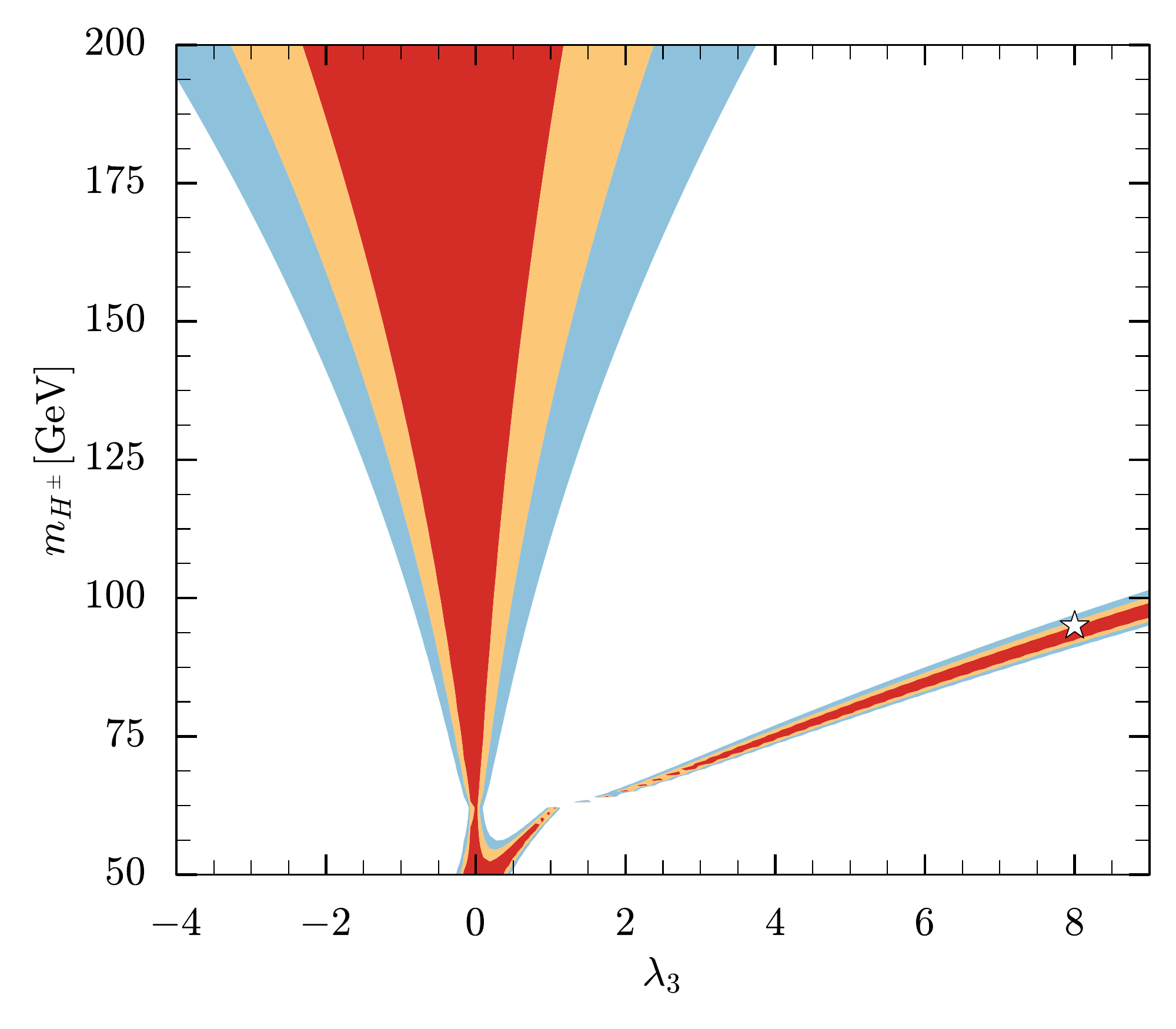}
\caption{Constraints on the model parameter space from oblique parameters in the mass of the charged Higgs mass and $\lambda_4$ (\emph{left}) and from Higgs signal strength measurements in the charged Higgs mass and $\lambda_4$ (\emph{right}). The red, yellow, and blue contours correspond to the $68\%$, $95\%$, and $99.7\%$ allowed regions respectively. Two different benchmark points corresponding to $(m_{H^\pm}, m_{A^0})=(95, 100) \ \mathrm{GeV}, (95, 160) \ \mathrm{GeV}$ are shown as black markers in the right panel. In the right panel, the white marker corresponds to the allowed value of $\lambda_3$ chosen in the rest of the analysis. Due to the constraints from neutrino mass and mixing parameters, we have taken $\lambda_5 = 0$.}
\label{exclusions:EWPT-gaga}
\end{figure}

Electroweak precision tests (EWPT) is a common approach to constrain
physics beyond SM by using the global fit through the oblique $S$, $T$ and $U$ parameters~\cite{Peskin:1991sw}. In the scotogenic model, the new gauge-inert
interactions will induce non-vanishing contributions to the oblique
parameters $\Delta T$ and $\Delta S$~\cite{Grimus:2008nb}\footnote{The corrections to the $U$-parameter in the IHDM are very small. Therefore, we assume that $\Delta U = 0$ in the present analysis.}. To study the impact of the EWPT on the mass splitting between the pseudo-scalar ($A^0$) and the charged Higgs boson ($H^\pm$), we have to minimize the function
\begin{eqnarray}
\chi^2 = \sum_{\mathcal{O}=S, T} \frac{(\mathcal{O} - \mathcal{O}_\mathrm{exp})^2}{\sigma_\mathcal{O}^2 (1- \rho_{ST}^2)} - 2 \rho_{ST} \frac{(S-S_\mathrm{exp})(T - T_\mathrm{exp})}{\sigma_S \sigma_T (1 - \rho_{ST}^2)},
\end{eqnarray}
with $S_\mathrm{exp}=0.06\pm 0.09, T_\mathrm{exp}=0.10\pm 0.07$ are the experimental values of the $S$ and the $T$ parameters, $\sigma_{S, T}$ are their corresponding errors, and $\rho_{ST} = + 0.91$ is their correlation. The constraints from EWPT can easily be satisfied in regions of the parameter space where the mass splitting between the neutral and the charged components of the inert doublet is small (for light scalars) or where the scalars are very heavy regardless the values of their mass splittings. In this model, constraints from neutrino masses and mixings imply extremely small values of $\lambda_5$. Therefore, the only parameter that is directly affected by EWPT constraint is $\lambda_4$. This is can easily be seen from the left panel of Fig. \ref{exclusions:EWPT-gaga} where we display the $1$-, $2$- and $3$-sigma allowed regions plotted in the $(m_{H^\pm}, \lambda_4)$ plane. We can see that, for e.g. $m_{H^\pm} \simeq 95 \ \mathrm{GeV}$, $\lambda_4$ can vary in $[-0.2, 1.5]$ which implies a maximum mass splitting of about $100$ GeV.

Moreover, the gauge bosons decay widths are well measured~\cite{Patrignani:2016xqp},
and must not be modified by any new interactions. Therefore, one needs
to impose the conditions $m_{H^{0}}+m_{A^{0}}, \, 2m_{H^{\pm}}>M_{Z};\, m_{H^{\pm}}+m_{A^{0}}, m_{H^{\pm}}+m_{H^{0}}>M_{W}, $ to keep the decay channels of
$W$ and $Z$ gauge bosons into inert particles closed.

The new Yukawa interactions in (\ref{LL}) lead to lepton flavor violating (LFV) decay processes that arise at one-loop level with
the exchange of charged Higgs $H^{\pm}$ and Majorana fermions $N_{k}$ particles.
The branching ratio of the decays $\ell_{\alpha}\rightarrow\ell_{\beta}+\gamma$
and $\ell_{\alpha}\rightarrow\ell_{\beta}\ell_{\beta}\ell_{\beta}$
are given in the literature~\cite{Toma:2013zsa}, and should be in
agreement with the available experimental constraints~\cite{Patrignani:2016xqp}.

In the scotogenic model, all the SM Higgs couplings with SM particles are the
same as in the SM except those relevant to the decays
$H\to \gamma \gamma$ and $H\to \gamma Z$ which receives additional
contributions from the charged Higgs bosons. Therefore, in the case where there is no large
contribution to the invisible decay of the SM Higgs, most of the LHC measurements would fit pretty well within the scotogenic model.
This is the case in our model, the only source of invisible decay is the one-loop induced coupling $H N_i N_j$ which is suppressed
in most regions of the parameter space. In the scotogenic model, the partial width of the SM Higgs boson in the $\gamma\gamma$ channel depends on the charged Higgs boson mass and $\lambda_3$. Positive (negative) values of the $\lambda_3$ would imply destructive (constructive) interferences with the leading $W$ and the sub-leading top quark contributions \cite{Arhrib:2015hoa}.
Since the charged Higgs $H^{\pm}$ contribution would modify the rate of
through $H\rightarrow\gamma\gamma$, we need to check the constraints on the parameter space from diphoton signal strength measurements at the LHC.
The public package \textsc{Lilith}
~\cite{Bernon:2015hsa, Kraml:2019sis} was used to check the constraints from various measurements of the Higgs boson signal strength ($\mu_{\gamma\gamma}^i$) defined by
\begin{eqnarray}
\mu_{\gamma\gamma}^i = \frac{\sigma^i \Gamma(H\to\gamma\gamma)}{\sigma^i_\mathrm{SM} \Gamma(H\to\gamma\gamma)^\mathrm{SM}},
\end{eqnarray}
with the superscript $i$ refers to the production channel of the SM Higgs boson. In the right panel of Fig. \ref{exclusions:EWPT-gaga}, we plot the allowed regions from Higgs boson signal strength measurements in the $(\lambda_3, m_{H^\pm})$ plane. We can see that $\mu_{\gamma\gamma}$ constrains strongly the 2D parameter space. There are two notable regions for $m_{H^\pm} < 125$ GeV; the first one is centered around $\lambda_3 \simeq 0$ while the second one is a small segment corresponding to $\lambda_3 \in [0, 9]$ and $m_{H^\pm} < 100$ GeV. This unsurprising because even for large positive values of $\lambda_3$, the $R_{\gamma\gamma}$ ratio in our model can still agree with data within the $+10\%$ of experimental uncertainty reported in e.g. the recent CMS analysis \cite{Sirunyan:2018koj}.

 In this study, we assume that the lightest right-handed Majorana neutrino
 is a DM candidate as was done in ref~\cite{Ahriche:2017iar}. For light Majorana neutrinos (with masses up to $140$ GeV) that we are interested
 in, the main annihilation channels are into charged leptons and SM neutrinos. These annihilation processes proceed through $t$-channel diagrams
 mediated by the members of the inert doublet. Furthermore, in the aim of simplifying the collider analysis (see sections~\ref{sec:LHC} and~\ref{sec:monoHiggs}), nearly degenerate
 Majorana neutrinos are chosen, i.e $m_{N_2} \simeq 1.01 m_{N_1}$ and $m_{N_3} \simeq 1.02 m_{N_1}$. In this case, co-annihilation becomes important and, therefore, is included
 in our analysis. Co-annihilation with inert scalars, which give rise to final states such as $\ell^\pm \gamma$, is sub-leading due to the smallness of the electromagnetic coupling compared to the new
 Yukawa couplings $h_{ik}$ and can be safely neglected. Including all the significant channels, we select a benchmark point that is in agreement with the \textsc{Wmap}~\cite{Hinshaw:2012aka} and \textsc{Planck}~\cite{Ade:2015xua}
measurements of the relic density at the $2\sigma$ level. \\

In our model, DM can interact with the nucleons and triggers a possible signal in direct
detection experiments. This can happen despite the absence of a tree level $H N_1 N_1$ coupling which arises
at the one-loop order. We estimated spin-independent (SI) scattering cross section of $N_1$ off a nucleon $\mathcal{N}$ and subject
it to constraints from the searches performed by \textsc{Xenon1T}~\cite{Aprile:2017iyp}. One notices that
constraints from direct detection are easily satisfied in our model due to the smallness of $H N_1 N_1$ coupling. We refer the reader to \cite{Ahriche:2017iar} for more details about the DM constraints in our model.

\subsection{LEP constraints}
Multiple searches for supersymmetric particles at $e^+ e^-$ collisions has been carried out by several collaborations~\cite{DELPHI:2003, Abbiendi:2003ji, Abbiendi:2003sc, Abdallah:2003xe} for
center-of-mass energies of $183$-$209$ GeV. The searches
focused on charginos and neutralinos pair production in events with two or three leptons and large transverse missing energy. Several interpretations in terms of models containing charged and neutral scalars have been made. Ref. \cite{Lundstrom:2008ai} made a comprehensive re-interpretation
of neutralino pair production ($\chi_1^0 \chi_2^0$) to constrain the production of $H^0A^0$ in
the IHDM and got a limit $m_{A^0} > 100$ GeV for large mass splitting. Pair production of charginos ($\chi_1^+ \chi_1^-$) was analyzed to put constraints on $H^\pm H^\mp$ production in a DM model with TeV scale colored particles \cite{Pierce:2007ut} and in the compressed IHDM~\cite{Blinov:2015qva}. \\

\begin{figure}[h]
\includegraphics[width=0.48\textwidth]{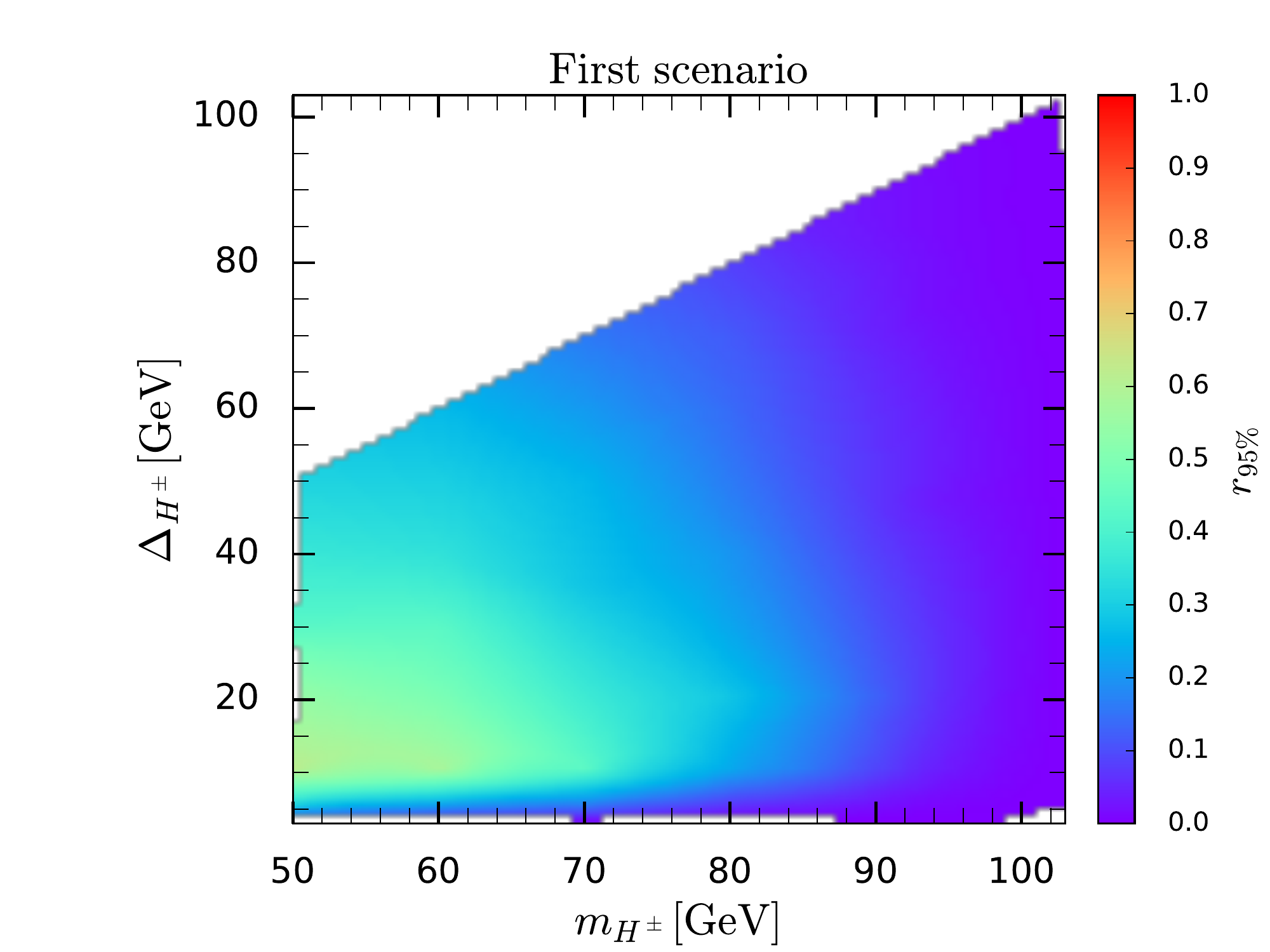}
\hfill
\includegraphics[width=0.48\textwidth]{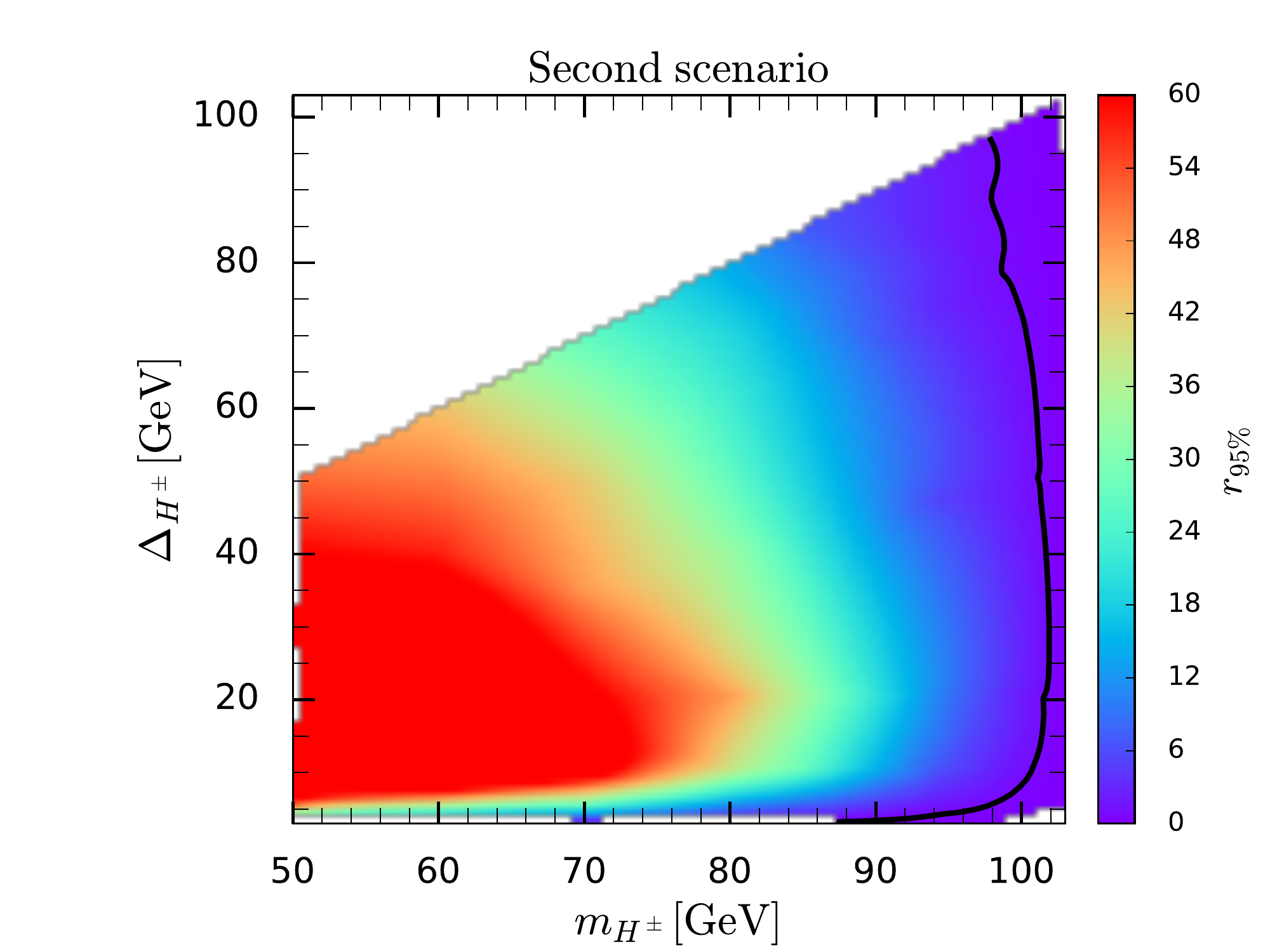}
\caption{Results of re-interpretation of LEP searches of chargino pair production on the ($\Delta_{H^\pm}, m_{H^\pm}$) plan in the first scenario
(\emph{left}) and second scenario (\emph{right}).}
\label{exclusions:LEP}
\end{figure}

In this section, we study the impact of LEP searches on the parameter space of our model. For instance, LEP put strong bounds on the pair production cross section of lightest neutralino. However, the LEP limits on neutralino pair production do not apply in the considered scenario of the scotogenic model, because the tiny value of the coupling $\lambda_5$ (of order $10^{-8}$-$10^{-10}$) required by the smallness of neutrino masses forbids off-shell decays, such as $A^0 \to H^0 Z \to H^0 \ell \ell$, and therefore yields an undetected final state. However, limits from chargino pair production can be applied to our model. Two processes can be used for such constraints; $e^+ e^- \to H^+ H^-$ and $e^+ e^- \to H^0 A^0$. The latter
contribute, if $\lambda_4 < 0$ and $\Delta_{H^\pm H^0} = m_{A^0, H^0} - m_{H^\pm} > m_{e, \mu}$, through off-shell decays. This contribution is proportional to $(\Delta_{H^\pm H^0})^5$ and, hence, is very small. Therefore, charged Higgs pair production is the only process to which the exclusion limits from LEP searches can be used to constrain our model. We consider the results of searches carried by
\textsc{Opal}~\cite{Abbiendi:2003sc} at $\sqrt{s}=208$ GeV and $\mathcal{L} = 680$ pb$^{-1}$ of integrated luminosity to derive conservative limits on the model parameter space i.e. by assuming that the efficiency of the selection is $100\%$\footnote{Full analysis of the signal process at the detector level will yield to an efficiency that is always smaller than $100\%$. Therefore, the limits we obtain in this study are more conservative.}

The pair production of the charged particle depends on new Yukawa couplings and the gauge couplings. The first contribution is proportional to
\begin{eqnarray}
\sum_{k=1}^{3} |h_{ek}|^2 = |h_{e1}|^2 + |h_{e2}|^2 + |h_{e3}|^2,
\end{eqnarray}
in the case of degenerate Majorana fermions. Because, for $\lambda_4 \geq 0$, the Charged Higgs boson decays with
$100\%$ branching ratio into $N_k \ell$, the limits from charginos searches can be used to constrain both the charged Higgs boson
and the mass splitting $\Delta_{H^\pm}$ defined by
\begin{eqnarray}
\Delta_{H^\pm} = m_{H^\pm} - m_{N_1}.
\end{eqnarray}
We consider two scenarios for the new Yukawa couplings; 1) where the Yukawa matrix is chosen as follows
\begin{eqnarray}
\frac{h_{ij}}{10^{-2}}=\left(\begin{array}{ccc}
-60.86-i0.20 & -0.30-i0.80 & 14.49-i0.75\\
25.14-i0.57 & -1.12-i2.49 & 40.87+i0.24\\
3.70+i0.62 &1.10+i3.88 & -44.20+i0.14
\end{array}\right)
\label{Yukawa}
\end{eqnarray}
which we called first scenario and 2) the second scenario where the $h_{ek}$ couplings take the highest values allowed by all the theoretical and experimental constraints ($h_{e1} = -0.026+i0.042 , h_{e2} = 2.22-i0.081, h_{e3} = 0.32-i0.0098$). In the second scenario,
the most important contribution comes from $|h_{e2}|$. \\
We estimate the $r_{95\%}$ ratio defined by

\begin{eqnarray}
r_{95\%} & = & \frac{\sigma(e^+ e^- \to H^+ H^-) \times (\textrm{BR}(H^\pm \to N_k \ell^\pm))^2}{95\% \sigma(e^+ e^- \to \chi_1^+ \chi_1^-) \times (\textrm{BR}(\chi_1^\pm \to \chi_1^0 \ell^\pm \nu_\ell))^2}\nonumber\\
 & = & \frac{\sigma(e^+ e^- \to H^+ H^-) }{95\% \sigma(e^+ e^- \to \chi_1^+ \chi_1^-) \times (\textrm{BR}(\chi_1^\pm \to \chi_1^0 \ell^\pm \nu_\ell))^2 },
 \label{LEP-r95}
\end{eqnarray}
where, in the second line of eq. (\ref{LEP-r95}), we used $\textrm{BR}(H^\pm \to \ell^\pm N_k) = 1$. A point in the parameter
space is excluded if the corresponding $r_{95\%}$ is larger than $1$. In Fig.~\ref{exclusions:LEP}, we depict the exclusions from
charginos pair production on the $(m_{H^\pm}, \Delta_{H^\pm})$ plan. As can be seen from the left panel of Fig.~\ref{exclusions:LEP},
all points are allowed by LEP searches. However, in the second scenario, one notices that the model is excluded for $m_{H^\pm} < 100$ GeV.
A small window corresponding to $\Delta_{H^\pm} < 5$ GeV and $90 \ \mathrm{GeV} < m_{H^\pm} < 100 \ \mathrm{GeV}$ is still allowed by these constraints.

\section{Constraints from LHC searches at $13$ TeV}
\label{sec:LHC}

The model parameter space can be constrained by re-interepreting several ATLAS and CMS searches for new physics beyond the SM. In this study,
we used the public tool \textsc{CheckMate}~\cite{Graesser:2012qy, Drees:2013wra, Buckley:2013kua, Dercks:2016npn, Kim:2015wza} which is
dedicated for re-interpretation of LHC searches of new physics. Degenerate Majorana neutrinos are chosen to avoid the possibility for
displaced vertices. The other parameters are fixed to avoid all the other theoretical and experimental constraints \cite{Ahriche:2017iar} and they are chosen to be
\begin{eqnarray}
\lambda_3 = 8, \quad m_{H^\pm} = 95 \ \textrm{GeV} \ \mathrm{and} \ m_{A^0} \in [100, 200] \ \mathrm{GeV}.
\end{eqnarray}
The LHC searches used in this analysis are displayed in Table~\ref{tab:constraints}. The model
parameter space can be affected by the LHC searches displayed in Table~\ref{tab:constraints} as we will show explicitly. Details about the
different searches performed at the LHC and the model-dependent processes that are sensitive to them are reported in~\ref{sec:appendix}.

In our model, new sources of missing transverse energy, $E_T^\textrm{miss}$, namely from right-handed neutrinos, $N_{i}$ exist. These
new sources can be probed at colliders with events triggered by large missing $E_T^\textrm{miss}$. However, Majorana neutrinos cannot
be produced directly because of the absence of the vertices $Z^{0}N\bar{N}$, $\gamma N\bar{N}$, and $H N\bar{N}$; right-handed neutrinos
are thus produced via the decays of the exotic scalars.

\begin{table}[h]
 \caption{Selected set of ATLAS and CMS searches that were used in the re-interpretation study. These analyses are implemented in \textsc{CheckMate}.}
 \label{tab:constraints}
 \begin{tabular*}{\textwidth}{@{\extracolsep{\fill}}rrrr@{}}
 \cline{1-4} \hline
 Analysis & Experiment & Luminosity ($\textrm{fb}^{-1}$) & Reference \\ \hline
 atlas\_conf\_2016\_050 & ATLAS & $13.3$ &~\cite{ATLAS:2016ljb} \\ \hline
 atlas\_conf\_2016\_066 & ATLAS & $13.3$ &~\cite{ATLAS:2016fks} \\ \hline
 atlas\_conf\_2016\_076 & ATLAS & $13.3$ &~\cite{ATLAS:2016xcm} \\ \hline
 atlas\_conf\_2017\_060 & ATLAS & $36.1$ &~\cite{ATLAS:2017dnw} \\ \hline
 atlas\_1704\_03848 & ATLAS & $36.1$ &~\cite{Aaboud:2017dor} \\ \hline
 atlas\_1709\_04183 & ATLAS & $36.1$ &~\cite{Aaboud:2017ayj} \\ \hline
 atlas\_1712\_02332 & ATLAS & $36.1$ &~\cite{Aaboud:2017vwy} \\ \hline
 atlas\_1712\_08119 & ATLAS & $36.1$ &~\cite{Aaboud:2017leg} \\ \hline
 atlas\_1802\_03158 & ATLAS & $36.1$ &~\cite{Aaboud:2018doq}
 \\ \hline
 cms\_sus\_16\_025 & CMS & $12.9$ &~\cite{CMS:2016zvj} \\ \hline
 cms\_sus\_16\_039 & CMS & $35.6$ &~\cite{Sirunyan:2017lae} \\ \hline
 cms\_sus\_16\_048 & CMS & $35.9$ &~\cite{Sirunyan:2018iwl} \\
 \cline{1-4} \hline
 \end{tabular*}
\end{table}

\begin{figure}[h]
\begin{center}
\includegraphics[width=0.6\textwidth]{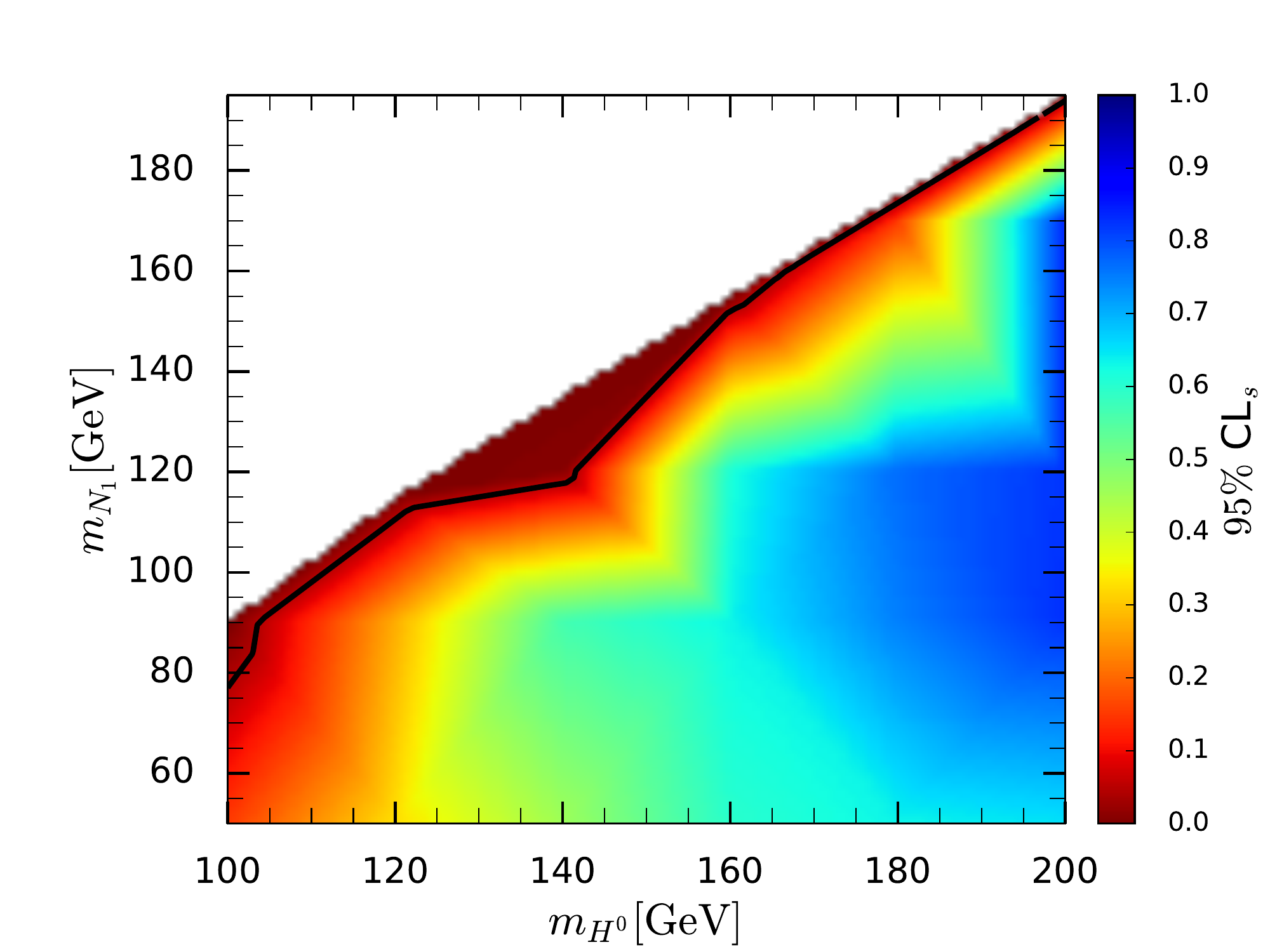}
\end{center}
\caption{Exclusions from LHC searches for new physics at $\sqrt{s}=13$ TeV projected on the ($m_{H^0}, m_{N_1}$) plan. The color map shows the
CL$_s$ values. The black line shows the excluded regions corresponding to CL$_s < 0.05$ while the white shaded area shows the region that is
forbidden by the constraint $m_{H^0} > m_{N_1}$.}\label{fig:LHC}
\end{figure}

In the degenerate window, since the decay $A^{0}\to H^{0}Z^{0}$ is kinematically forbidden, the scalar/pseuodoscalar can be produced in association with a charged scalar which subsequently decays to a charged lepton and a right-handed neutrino. While the scalar and pseudoscalar may only decay invisibly; we obtain a signal with a single lepton and large missing $E_T^\textrm{miss}$. In this channel the most sensitive LHC search comes from the work in~\cite{ATLAS:2016ljb} that searches for SUSY in a final state with one isolated lepton. In the case where the exotic scalars are pair produced, in the degenerate region, their decays lead only to missing $E_{T}^\textrm{miss}$ and one can tag this channel with a mono-jet from initial state radiation. In these cases, LHC searches with photons and jets are the most sensitive, with the largest amount of missing $E_{T}^\textrm{miss}$ when the scalar/pseudoscalar mass approaches the right-handed neutrino mass, and this is where the bulk of the exclusion lies in as can be seen from Figure~\ref{fig:LHC} after the inclusion of all relevant LHC searches given in Table~\ref{tab:constraints}. Following the results of the re-interpretation of LHC searches of new physics that we have shown in Fig.~\ref{fig:LHC}, we choose the following benchmark points for the mono-Higgs study;

\begin{eqnarray}
100~\textrm{GeV} \leq m_{H^0} & = & m_{A^0} \leq 200~\textrm{GeV}, ~m_{H^\pm} = 95 \ \textrm{GeV} \nonumber\\
\quad m_{N_1} = m_{N_2} & = & m_{N_3} = 80~\textrm{GeV}, ~\lambda_3 = 8.
\end{eqnarray}
while the new Yukawa couplings are fixed to their values shown in eq.(\ref{Yukawa}).

\section{Mono-Higgs signature}
\label{sec:monoHiggs}
In this section, we describe different aspects of our analysis. First we discuss the contribution to the signal process as well as the possible backgrounds and the corresponding cross sections. Then, we
discuss in depth the phenomenological setup used in our analysis and event selection.
\subsection{Signal and backgrounds}
In this model, mono-Higgs production proceeds through two different processes, i.e
\begin{eqnarray}
p p \to S S H \to N_i N_j \nu \bar{\nu} H
\end{eqnarray}
and
\begin{eqnarray}
p p \to N_i N_j H.
\end{eqnarray}
\begin{figure}[h]
\begin{center}
\includegraphics[width=0.75\textwidth]{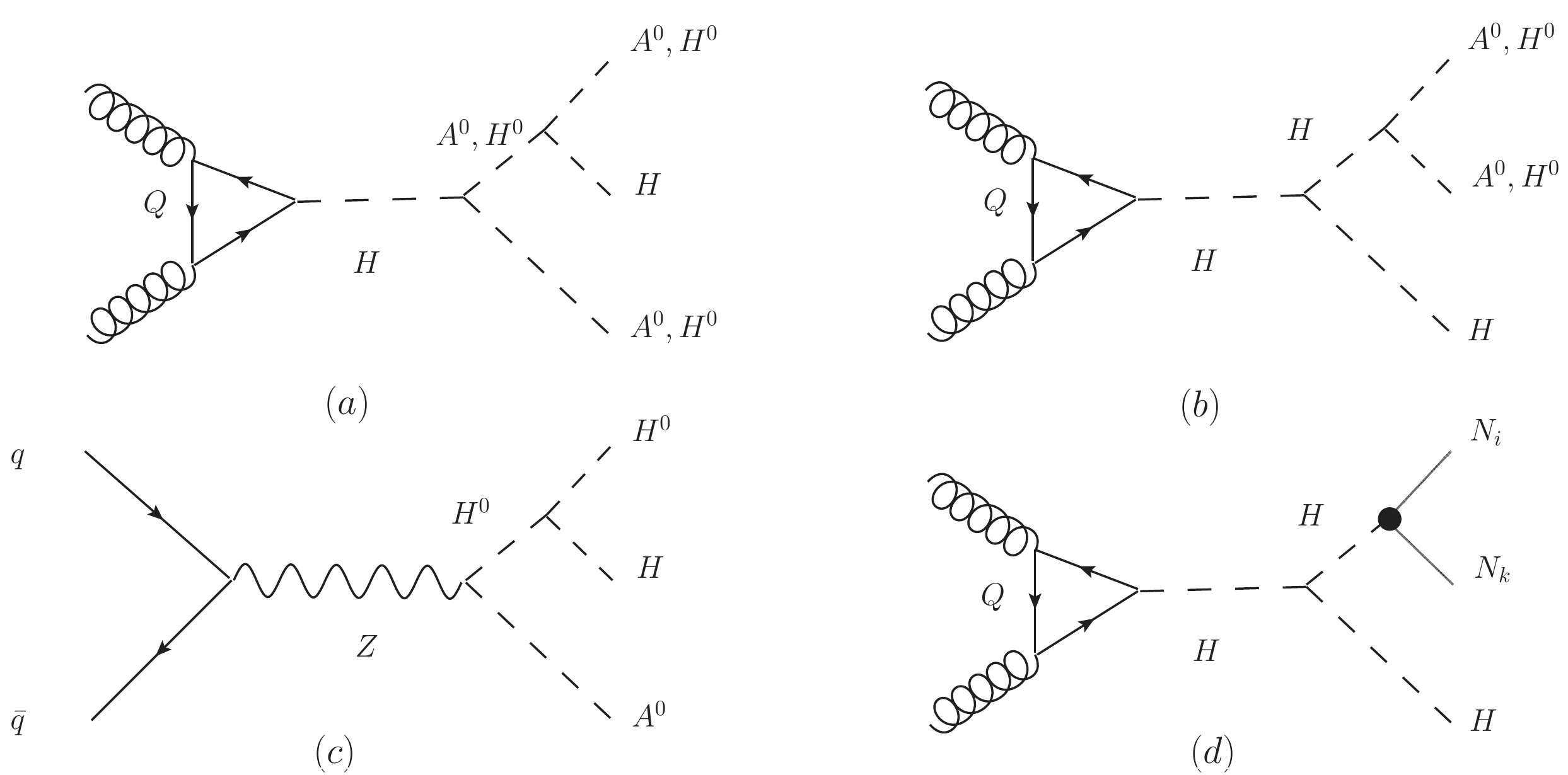}
\end{center}
\caption{Parton level Feynman diagrams contributing to the mono-Higgs signal in hadronic collisions. Unlike the fourth diagram, the first three diagrams are
efficient only when the decays $H^0/A^0\rightarrow W^{\pm}H^{\mp}$ have extremely small branching fractions. }
\label{fig:diagrams}
\end{figure}

The corresponding Feynman diagrams are depicted in Fig.~\ref{fig:diagrams}. There are four contributions to Higgs+$E_T^\textrm{miss}$ signal in hadronic collisions which involve either the production of an off-shell Higgs boson or a $Z$-boson. In the first diagram (\ref{fig:diagrams}-a), the off-shell Higgs boson splits into $SSH$ while in the second one, it involves a contribution from the SM Higgs trilinear coupling $\lambda_{HHH}$ (\ref{fig:diagrams}-b). In the third contribution (\ref{fig:diagrams}-c), $q\bar{q}$ annihilates into a $Z^*$ which splits into two dark Higgses. The fourth contribution consists of two Majorana neutrinos produced in association with a SM Higgs boson (\ref{fig:diagrams}-d). The first and second contributions interfere destructively (constructively) for negative (positive) values of the $HSS$ couplings. We notice that the contribution of diagram-c is the most dominant one as it contributes about $95\%$ of the total cross section. This is unsurprising since this contribution occurs at the tree level and is enhanced for large values of $\lambda_L$. Using simple power counting, one notices that the total cross section behaves as
\begin{eqnarray}
\sigma & \propto & \bigg| \lambda_L^2 \mathcal{M}_a + \lambda_L \lambda_{HHH} \mathcal{M}_b \bigg|^2 + \bigg|\lambda_L \mathcal{M}_c \bigg|^2 + \bigg|\sum_{i, j=1}^3 \tilde{y}_{HN_iN_j} \lambda_{HHH} \mathcal{M}_d \bigg|^2.
\label{monoHiggs-behav}
\end{eqnarray}

\begin{figure}[h]
\begin{center}
\includegraphics[width=0.6\textwidth]{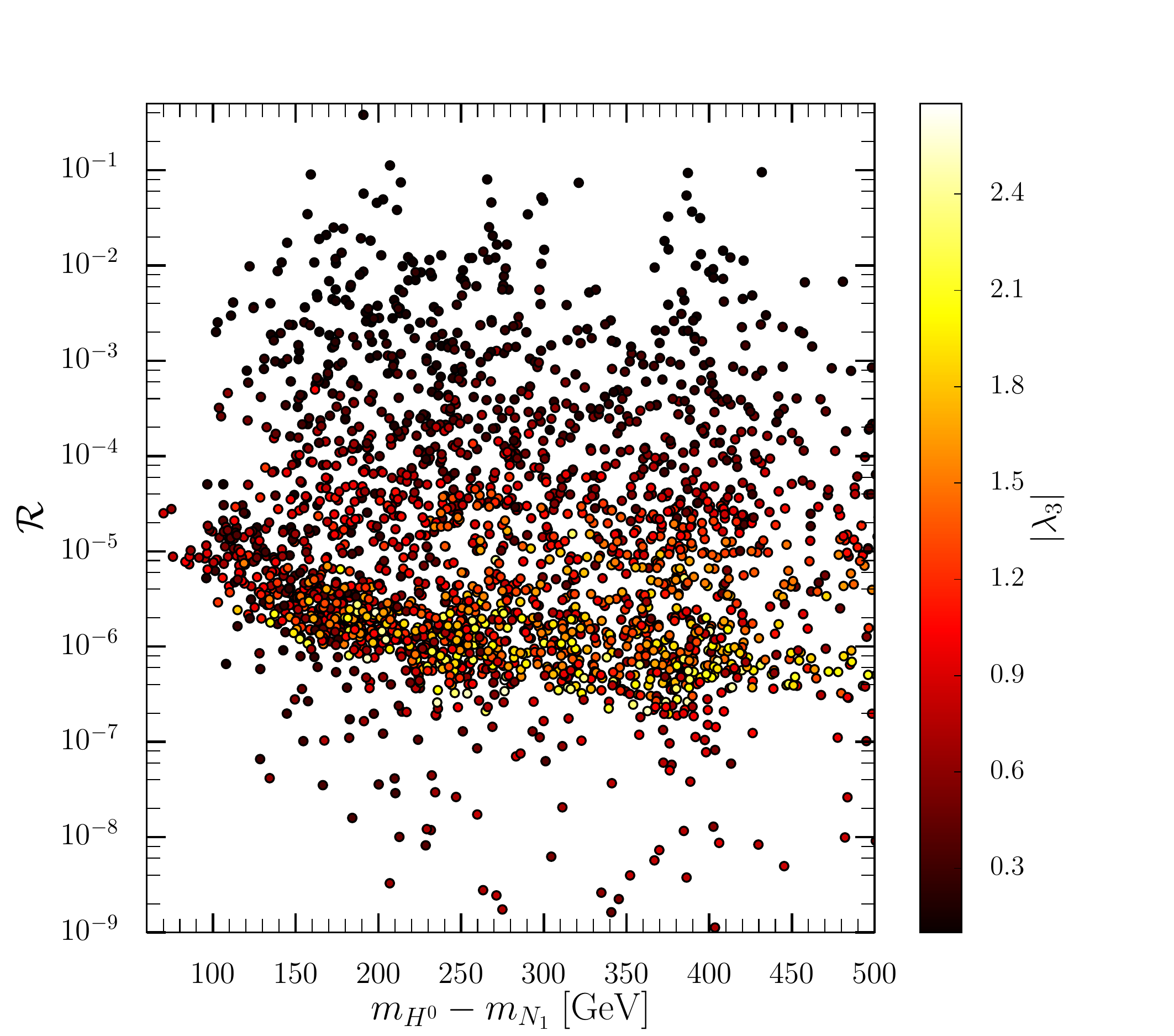}
\end{center}
\caption{$\mathcal{R}$, defined in eq. (\ref{ratio}), as a function of $m_{H^0}-m_{N_1}$. The color map shows the values of $|\lambda_L|$.
 The points shown in the plot satisfy all the theoretical and experimental constraints discussed in section~\ref{sec:model}.}
\label{fig:ratio}
\end{figure}

\begin{figure}[h]
\includegraphics[width=0.48\textwidth]{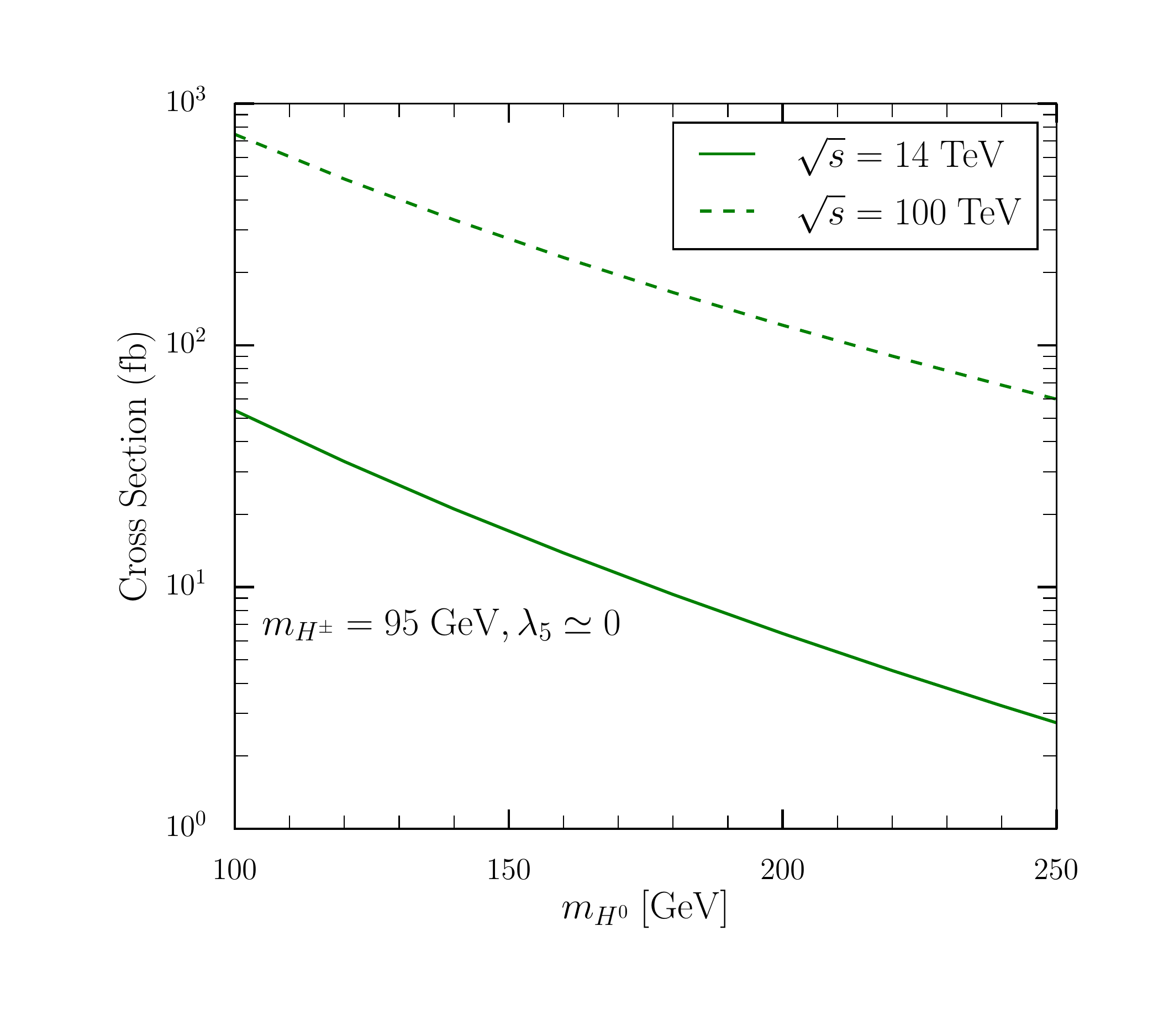}
\hfill
\includegraphics[width=0.48\textwidth]{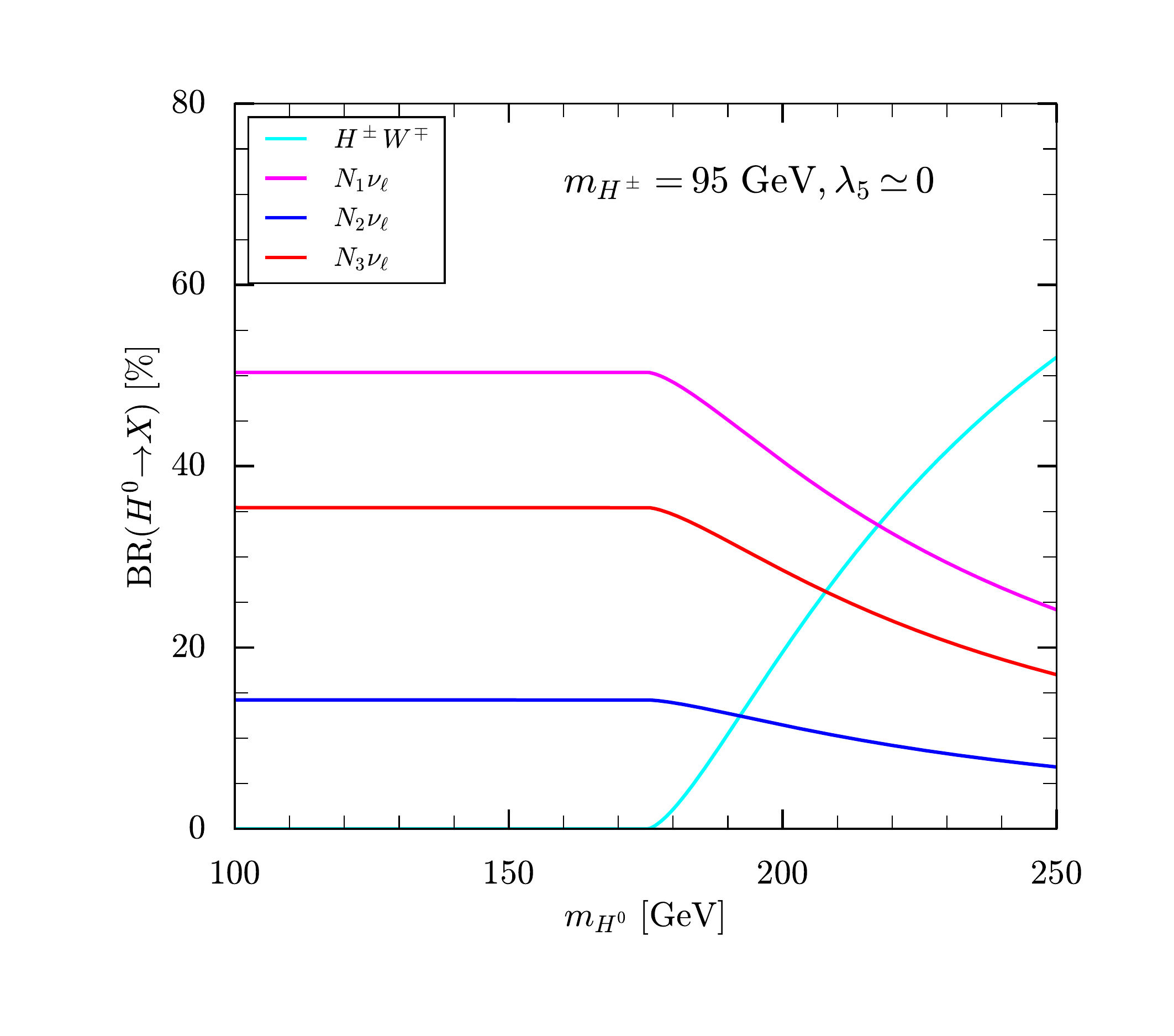}
\caption{\emph{Left:} Mono-Higgs boson production cross section as a function of the Dark scalar mass $m_{H^0}\simeq m_{A^0}$ for $m_{H^\pm}=95 \ \mathrm{GeV}$ at the LHC (solid line) and at a future $100$ TeV collider (dashed). We included the processes $gg\to H^0 H^0 H, gg\to A^0 A^0 H$ and $q\bar{q}\to H^0 A^0 H$. The depicted results were computed LO with \textsc{Madgraph5\_aMC@NLO}. \emph{Right:} Decay branching ratios of the Dark scalar particle as a function of the Dark scalar mass.}
\label{fig:monHiggs-xsection}
\end{figure}

The contribution of diagram~\ref{fig:diagrams}-d is proportional to the squared of the $HN_iN_j$ coupling which is one-loop
induced~\cite{Ahriche:2017iar} and it is expected to be very small. In this regard, we define the ratio $\mathcal{R}$ by
\begin{eqnarray}
\mathcal{R} = \frac{\sum_{i, j=1}^3 |\tilde{y}_{HN_iN_j}|^2}{|\lambda_L|^4},
\label{ratio}
\end{eqnarray}
which gives a rough estimate of the relative contribution of diagram
~\ref{fig:diagrams}-d to the signal cross section where only the leading contribution to $SSH$ production ($\simeq |\lambda_L|^4$) is included.
We show this ratio in Fig.~\ref{fig:ratio} as function of the mass splitting $\Delta m_{NH^0}=m_{H^0}-m_{N_k}$ with a color map showing $|\lambda_3|$.
One can see that this ratio can only be important for very small values of $\lambda_L$, i.e $|\lambda_L| < 0.1$. Given that this region is not
interesting from phenomenological point of view as it yields very small cross sections (see Fig.~\ref{fig:monHiggs-xsection}), we conclude that the
contribution of diagram (\ref{fig:diagrams}-d) can be safely neglected.

The cross sections for the mono-Higgs production are depicted in the left panel of Fig.~\ref{fig:monHiggs-xsection}. As expected, one can see that the cross
section is pretty small for the LHC at $\sqrt{s}=14$ TeV with the maximum being $\sigma_\textrm{max} \simeq 53$ fb for $m_{H^0} = 100$ GeV which increases by about an order of magnitude at the FCC-hh with 100 TeV. Since the mass splitting $\Delta_{H^\pm H^0}$ can be as large as $100$ GeV, the dark neutral (pseudo)-scalar does not always decay exclusively into an invisible final state. Therefore, in order to estimate correctly the number of events in a signal benchmark point, one has to scale correctly the corresponding cross section by $\textrm{BR}(H^0 \to \textrm{invisible})^2$. We show the Dark scalar branching ratios as a function of $m_{H^0}$ in Fig. \ref{fig:monHiggs-xsection} (right). We can see that, unless $m_{H^0} > 190 \ \textrm{GeV}$, the invisible decays of $H^0$ have always a branching fraction larger than $90\%$.

\begin{table}[h]
\caption{Cross sections for processes contributing to the Higgs+$E_T^\textrm{miss}$ background. The numbers outside (inside) the
brackets refers to the rates at $14$ ($100$) TeV. Details about the computation are explained in the text. Here, $\sigma\times \textrm{BR}$
refers to $\sigma(gg\to H) \times \textrm{BR}(H\to \gamma\gamma)$ for $gg\to H$, and to $\sigma(pp \to ZH)\times \textrm{BR}(H\to\gamma\gamma)\times \textrm{BR}(Z\to \bar{\nu_\ell} \nu_\ell)$ for $ZH$, to $\sigma(pp \to W^\pm H)\times \textrm{BR}(H\to\gamma\gamma)\times \textrm{BR}(W^\pm\to \ell^\pm\nu_\ell)$ for the case of $W^\pm H$ and to $\sigma(pp \to W^\pm + n \gamma) \times \textrm{BR}(W^\pm\to \ell^\pm\nu_\ell) + \sigma(pp \to Z + n \gamma)\times \textrm{BR}(Z \to \bar{\nu}_\ell \nu_\ell)$ for $V + n\gamma, n=1, 2$.}
 \label{background-Higgs}
 \begin{tabular*}{\textwidth}{@{\extracolsep{\fill}}rrrr@{}}
 \cline{1-4} \hline
 \hspace{0.5cm} Process \hspace{0.5cm} & $\sigma\times \textrm{BR}$ [fb] & Generator & Perturbative Order \\
 \cline{1-4} \hline
 $gg \to H$ & $128.54~(1.94 \times 10^3)$ & \textsc{Sushi}~\cite{Harlander:2012pb, Harlander:2016hcx} & NNNLO \\ \hline
 $W^\pm H$ & $1.16~(12.59)$ & \textsc{Vh@nnlo}~\cite{Brein:2012ne} & NNLO \\ \hline
 $Z H$ & $0.52~(7.34)$ & \textsc{Vh@nnlo}~\cite{Brein:2012ne} & NNLO \\ \hline
 $V\gamma\gamma$ & $51.99~(621.96)$ & \textsc{Madgraph5\_aMC@NLO}~\cite{Alwall:2014hca} & NLO \\ \hline
 $V\gamma$ & $42.89\times10^3~(397.04\times10^3)$ & \textsc{Madgraph5\_aMC@NLO}~\cite{Alwall:2014hca} & NLO\\ \hline
 $\gamma\gamma$+jets & $4.19 \times 10^6~(52.81 \times 10^6)$ & \textsc{Sherpa}~\cite{Gleisberg:2008ta} & NLO \\ \cline{1-4} \hline
 \end{tabular*}
\end{table}


The $\gamma\gamma$ decay channel represents a very clean signature of the mono-Higgs final state boson despite the smallness of
the corresponding branching ratio (which is about $\simeq0.23\%$). In this case, the following backgrounds have to be considered
 \begin{itemize}
 \item $gg \to H \to \gamma\gamma$: this is the dominant background. The missing energy is due to the mis-identification of soft QCD radiation. However, it can be substantially suppressed by requiring high missing transverse energy as we will show later on.
 \item $p p \to Z H$: where the $Z$-boson decays to a pair of neutrinos is an irreducible background. The suppression of this background can be achieved by applying specific selection criteria, e.g on the transverse mass of the (Higgs, $E_T^\textrm{miss}$) system.
 \item $pp \to W^\pm H$: where the $W^\pm$-boson decays into $\ell^\pm \nu$ where the charged lepton escapes the detection, i.e not passing the selection threshold. At the LHC, the charged lepton efficiency is high and, therefore, we expect that this background will have small contribution.
 \item $pp \to V\gamma\gamma$: where the $V=Z$-boson decays invisibly and the $V=W$-boson decays leptonically. The $Z\gamma\gamma$ background is irreducible contrarily to the $W\gamma\gamma$. The contribution of the latter can be reduced by imposing a lepton veto in the selection procedure. Both the two backgrounds have weaker $\gamma$ spectrum and, therefore, their contribution can be weakened by strong requirements on the $p_T^\gamma$ and the invariant mass of the $\gamma\gamma$ spectrum.
\item $pp \to V\gamma$: this background is similar to $V\gamma\gamma$.
 \item $pp \to \gamma\gamma+$jets: In the hadronic environment, there is a possibility that pile-up events will contributes to fake high missing transverse energy. The rate of this process is very high and we opt to generate parton level cross sections with some cuts on the $p_T$ of photons and jets. ATLAS~\cite{Aaboud:2017uak} and CMS~\cite{Sirunyan:2018fpy} collaborations used different strategies to reduce the contribution of this background either by defining some kinematical variables or use azimuthal separation between the reconstructed Higgs candidate and the missing transverse energy. These features will be discussed briefly in the next subsection.
 \end{itemize}

\begin{figure}[h]
\includegraphics[width=0.48\textwidth, height=0.38\textwidth]{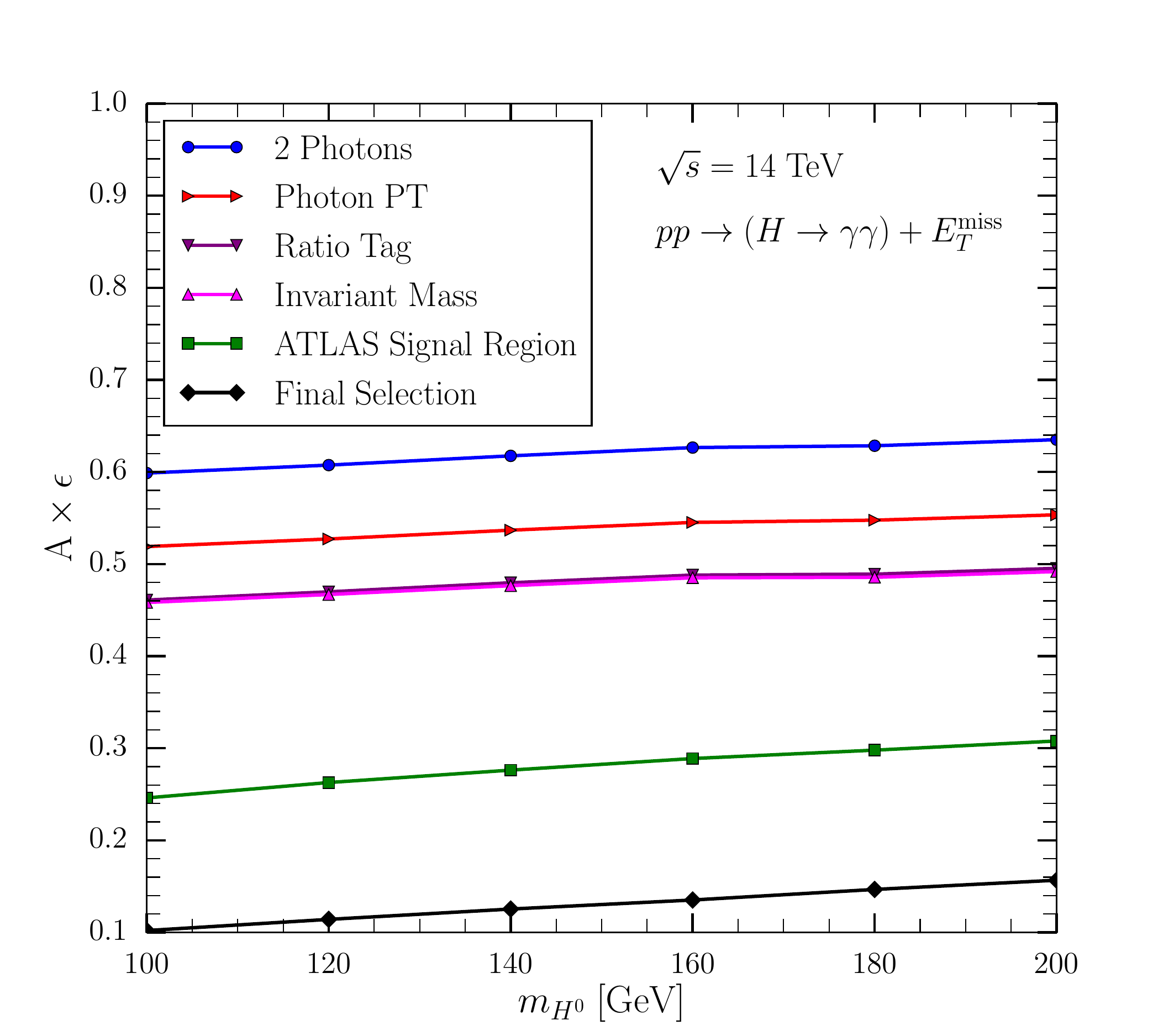}
\hfill
\includegraphics[width=0.48\textwidth, height=0.38\textwidth]{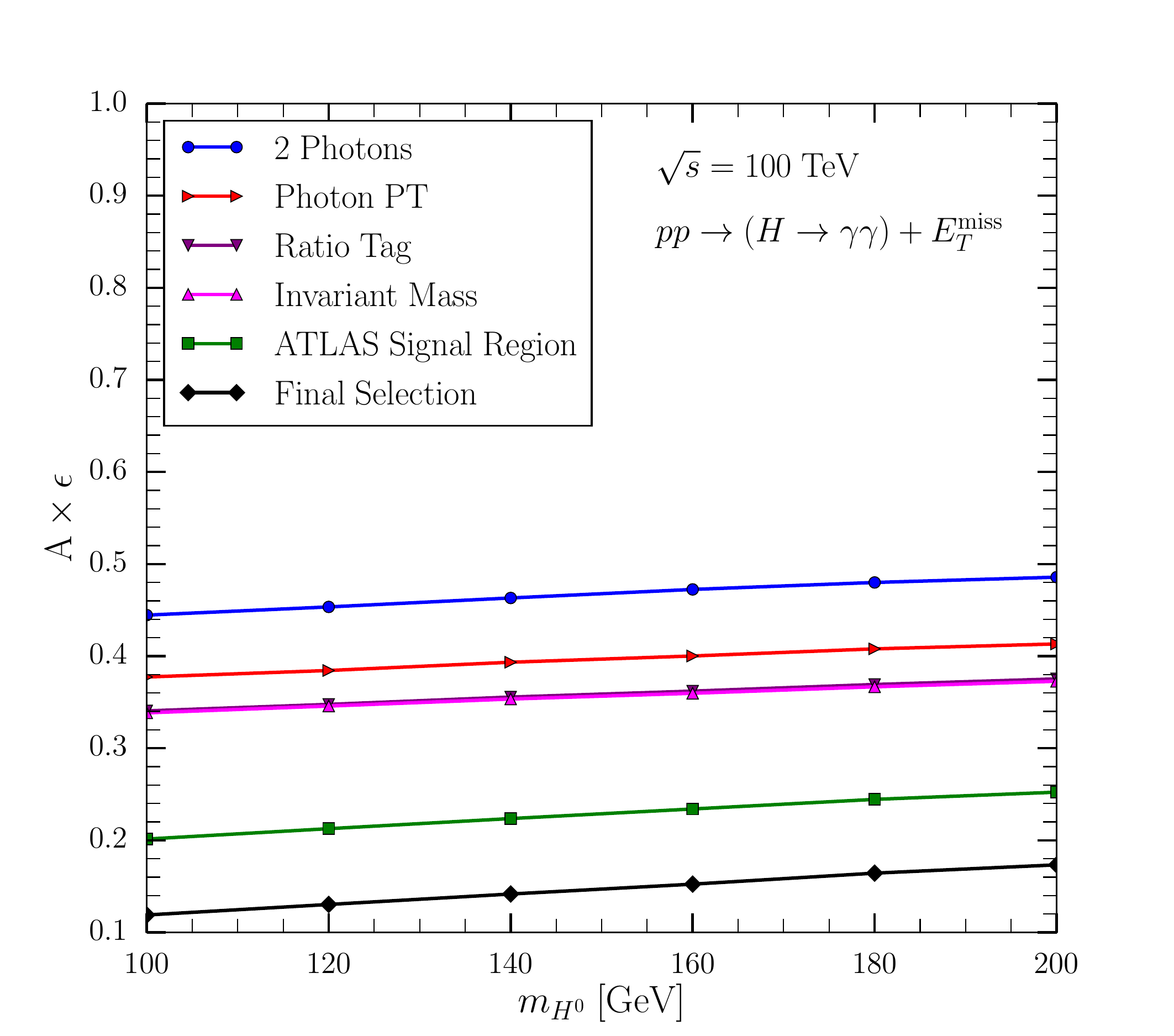}
\caption{The acceptance times the efficiency ($A\times\epsilon$) for the signal after each step of the event selection as a function of the
dark Higgs mass for $\sqrt{s}=14$ TeV (\emph{left}) and $\sqrt{s}=100$ TeV (\emph{right}). We show $A\times\epsilon$ after the "2 Photons" selection step
(blue), for the events passing the "Photon PT" selection (red), after the photon isolation selection denoted by "Ratio Tag" (purple) and for events in
which the invariant mass of the diphoton system falls in the interval $m_{\gamma\gamma} \in [110, 160]~\textrm{GeV}$ (rose). The efficiency for the
two signal regions are shown in green (ATLAS signal region) and in black (tight selection).}
\label{fig:efficiency}
\end{figure}

\begin{table}[h]
 \caption{Cut flow for $H\to\gamma\gamma$ final state at the LHC at $\sqrt{s}=14$ TeV and for $3$ ab$^{-1}$ of luminosity.}
 \label{tab:cutflow1}
 \begin{tabular*}{\textwidth}{@{\extracolsep{\fill}}rrrrrr@{}}
 \cline{1-6} \hline
 Cuts & SM Higgs & $V\gamma\gamma, V\gamma$ & $\gamma\gamma+$jets & Signal & $S/B$ \\ \hline
 Initial events & $322359$ & $128432167$ & $24030000$ & $365$ & $2.4 \times 10^{-6}$ \\ \hline
 2 Photons & $168005$ & $2548352$ & $6837913$ & $218$ & $2.3 \times 10^{-5}$ \\ \hline
 Photon PT & $150570$ & $1177335$ & $6317283$ & $189$ & $2.5 \times 10^{-5}$ \\ \hline
 Ratio Tag & $135720$ & $830147$ & $5582001$ & $168$ & $2.6 \times 10^{-5}$ \\ \hline
 Invariant Mass & $135492$ & $174358$ & $2066511$ & $166$ & $6.9 \times 10^{-5}$ \\ \hline
 ATLAS Signal Region & $98$ & $151$ & $0$ & $89$ & $0.35$ \\ \hline
 Final Selection & $29$ & $5$ & $0$ & $32$ & $0.94$ \\
 \cline{1-6} \hline
 \end{tabular*}
\end{table}

\begin{table}[h]
 \caption{Cut flow for $H\to\gamma\gamma$ final state at the LHC at $\sqrt{s}=100$ TeV and for $3$ ab$^{-1}$ of luminosity.}
 \label{tab:cutflow2}
 \begin{tabular*}{\textwidth}{@{\extracolsep{\fill}}rrrrrr@{}}
 \cline{1-6} \hline
 Cuts & SM Higgs & $V\gamma\gamma, V\gamma$ & $\gamma\gamma+$jets & Signal & $S/B$ \\ \hline
 Initial events & $5885817$ & $1192995664$ & $252090000$ & $5147$ & $3.5 \times 10^{-6}$ \\ \hline
 2 Photons & $2597337$ & $18298491$ & $73202377$ & $2287$ & $2.4 \times 10^{-5}$ \\ \hline
 Photon PT & $2272845$ & $8405387$ & $67325991$ & $1941$ & $2.5 \times 10^{-5}$ \\ \hline
 Ratio Tag & $2051298$ & $6134810$ & $59189681$ & $1753$ & $2.6 \times 10^{-5}$ \\ \hline
 Invariant Mass & $2048497$ & $1228567$ & $21714801$ & $1741$ & $6.9 \times 10^{-5}$ \\ \hline
 ATLAS Signal Region & $4882$ & $1889$ & $0$ & $1036$ & $0.15$ \\ \hline
 Final Selection & $2215$ & $315$ & $0$ & $612$ & $0.24$ \\
 \cline{1-6} \hline
 \end{tabular*}
\end{table}

\subsection{Phenomenological setup and Event selection}
The cross sections of the background processes are depicted in Table~\ref{background-Higgs} for both the LHC at $\sqrt{s}=14$ TeV and FCC-hh
at $\sqrt{s}=100$ TeV. The cross section of $gg \to H$ was computed at NNNLO using \textsc{Sushi}~\cite{Harlander:2012pb, Harlander:2016hcx}
version 1.6.1 which implements the results of~\cite{Chetyrkin:2000yt, Harlander:2002wh, Anastasiou:2014lda, Anastasiou:2015yha, Anastasiou:2016cez}.
The rates for $W^\pm H$ and $Z H$ processes were estimated at NNLO~\cite{Harlander:2002wh, Brein:2003wg} inlcuding NLO EW corrections~\cite{Ciccolini:2003jy}
and top quark mass effects~\cite{Brein:2011vx} using the public package \textsc{Vh@nnlo}~\cite{Brein:2012ne} version 2.0.3. In all the NNLO calculations,
the \textsc{CT10} PDF set~\cite{Gao:2013xoa} was used with $\alpha_s(M_Z^2)=0.118$. The cross section for $
 V\gamma$ and $V\gamma\gamma$ was evaluated at NLO using \textsc{Madgraph5\_aMC@NLO}~\cite{Alwall:2014hca} with the \textsc{Nnpdf30} PDF sets~\cite{Ball:2014uwa}.
The estimate of $\gamma\gamma$ process (excluding $H$ contribution) was done using \textsc{Sherpa} version 2.2.5~\cite{Gleisberg:2008ta} where inclusive
samples of multipliticity up to $4$ jets in the final state are merged using the CKKW matching scheme~\cite{Catani:2001cc} and a merging scale $Q_0 = 20$ GeV. \\

Events for both the signal and the backgrounds were generated using \textsc{Madgraph5\_aMC@NLO} and \\ \textsc{Pythia8}~\cite{Sjostrand:2014zea} at LO in QCD. Background events involving the Higgs boson were generated and decayed with \textsc{Pythia8} while $V\gamma$ and $V\gamma\gamma$ events were generated using \textsc{Madgraph5\_aMC@NLO} including the leptonic decays of the massive electroweak gauge bosons. The $\gamma\gamma$+jets events were generated with \textsc{Pythia} and normalized to their rate at NLO. Since the rate of this process is huge, and most of the events will be vetoed in the initial selection, events are generated with a $p_{T, \gamma}^\textrm{min} = 70$ GeV, and $|\eta^\gamma| < 2.5$. Events for $gg\to H$ were scaled by a $K$-factor of $3.2$ using the results of \textsc{Sushi} while $VH$ events were scaled by a factor of $1.6$. All the background events were showered with \textsc{Pythia}. \textsc{Delphes3} was used for fast detector simulation~\cite{deFavereau:2013fsa}. \\

The analysis of events was carried out at the detector level using implemented efficiencies, and mis-identification rates in \textsc{Delphes}
where the parameters are tuned for the ATLAS experiment and extrapolated for a future FCC-hh~\cite{Mangano:2017tke}. Events pass a preselection
stage with all the objects (leptons, jets, photons and missing $E_T$) are kept. The Acceptance times the efficiency ($A\times \epsilon$) is
depicted in Fig.~\ref{fig:efficiency} as function of $m_{H^0}$ for $\sqrt{s}=14$ TeV and $\sqrt{s}=100$ TeV. The cutflow for the event
selection is shown in Tables~\ref{tab:cutflow1} and~\ref{tab:cutflow2}. Events are selected if they contain at least two photons with
$p_T^\gamma > 25$ GeV and $|\eta^\gamma| < 2.37$. This selection is denoted by "Photon PT" in Fig.~\ref{fig:efficiency} and
Tables~\ref{tab:cutflow1} and~\ref{tab:cutflow2}. Besides, we do not impose any requirement on the multiplicity, hardness and flavor
compositions of jets or the multiplicity of charged leptons. The photons that pass the initial selection will be subject to further
isolation cuts (as in~\cite{Aaboud:2017uak}), and the photon candidates are ordered by their transverse momentum.
The two leading photons are used to reconstruct a \emph{Higgs candidate}. Further, The ratio of the transverse momentum to the
invariant mass $p_T^\gamma/m_{\gamma\gamma}$ is required to be larger than $0.35$ ($0.25$) for the leading (sub-leading) photon.
Furthermore, a cut on the invariant mass of the diphoton system is imposed; namely events are selected if
$110~\textrm{GeV} < m_{\gamma\gamma} < 160~\textrm{GeV}$. But in some cases, events in the $\gamma\gamma$ and $\gamma+$jets
backgrounds contain large fake transverse missing energy, which is due to the fact that, in such events, the vertex with larger
$\sum p_T^2$ (where the sum runs over all the tracks) is not the primary vertex but the one coming from pile-up.\footnote{A primary
vertex is defined as the spatial point where proton-proton collisions occur.} Both ATLAS and CMS collaborations used sophisticated methods
to reduce the contribution of pile-up to missing transverse energy. The ATLAS collaboration has defined a new variable $S_{E_T^\textrm{miss}}$ defined by
\begin{eqnarray}
S_{E_T^\textrm{miss}} = \frac{E_T^\textrm{miss}}{\sqrt{\sum_i E_T^i}},
\label{SEtmiss}
\end{eqnarray}
where $i$ correspond to all the objects (photons, jets, and leptons) used to construct the missing transverse energy. Besides, to improve the resolution of the $E_T^\textrm{miss}$, tracks and clusters not associated to the diphoton primary vertex are not used to reconstruct $E_T^\textrm{miss}$~\cite{Aaboud:2017uak}. The CMS collaboration used variables that characterize the back-to-back event topology of the signal events (for instance $|\Delta\Phi(E_T^\textrm{miss}, \vec{p}_{\gamma\gamma})|$). By requiring that such quantity is larger than $2.1$, only events where the reconstructed Higgs and missing transverse energy are back-to-back are selected. Therefore, the contribution from e.g. $\gamma\gamma$ backgrounds is significantly reduced.

We compared between the approaches used by ATLAS and CMS to reduce the contribution from $\gamma\gamma$+jets backgrounds, on our benchmark points; and we found that they produce results that agree with each other. We will follow the ATLAS selection criteria throughout this study. We define two signal regions; the mono-Higgs signal region (denoted by ATLAS signal region in this paper) and a tight signal region. The kinematical quantities and selection rules are displayed in Table~\ref{tab:signalregion}. In the two signal regions, we require that the invariant mass of the diphoton system falls inside the interval $[115, 135]~\textrm{GeV}$. Furthermore, we require that events contain no lepton (either electron or muon) with $p_T^\ell > 10$ GeV and $|\eta^\ell| < 2.5$.

\begin{table}[h]
 \caption{Selection rule used to enhance the significance for $H(\to\gamma\gamma)+E_T^\textrm{miss}$ final state.}
 \label{tab:signalregion}
 \begin{tabular*}{\textwidth}{@{\extracolsep{\fill}}rr@{}}
 \cline{1-2} \hline
 Signal region & Cuts \\ \hline
 ATLAS signal region & $p_T^{\gamma\gamma} > 90~\textrm{GeV}$, $S_{E_T^\textrm{miss}} > 7$. \\ \hline
 Tight selection & $p_T^{\gamma\gamma} > 90~\textrm{GeV}$, $S_{E_T^\textrm{miss}} > 7$, \\
 & $E_T^\textrm{miss} > 200~\textrm{GeV}$, $p_T^\gamma(\textrm{lead}) > 40~\textrm{GeV}$ \\\cline{1-2} \hline
 \end{tabular*}
\end{table}

At $\sqrt{s} = 14$ TeV, we can see from Table~\ref{tab:cutflow1} that the signal-to-background ratio ($S/B$) can go from $\simeq 10^{-5}$ (after the first selection) to about $\simeq 1$ in the mono-Higgs signal region. Besides, the efficiency of the signal for $m_{H^0}=100$ GeV is $A\times \epsilon\simeq 25\%$ in the ATLAS signal region. For the FCC-hh at $100$ TeV, the signal-to-background ratio can go up to $\simeq 0.24$ in the tight signal region. If one requires, in addition to the tight selection rules, that $p_T^\gamma > 60$ GeV (for the leading photon) and $p_T^\gamma > 50$ GeV (for the sub-leading photon), the significance can increase to around $\simeq 20$ but the statistics goes down by about an order of magnitude.

\begin{figure}[h]
\includegraphics[width=0.48\textwidth, height=0.38\textwidth]{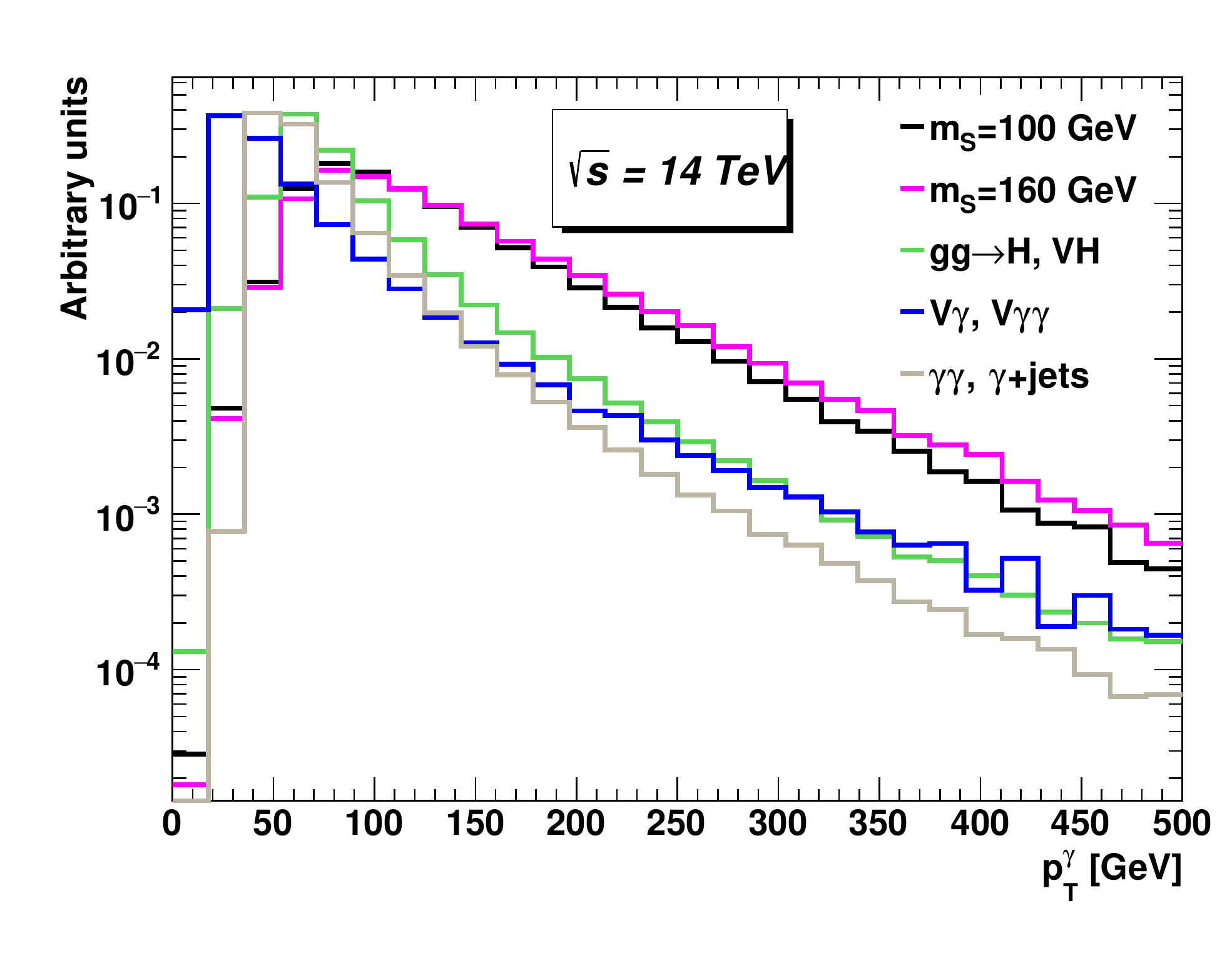}
\hfill
\includegraphics[width=0.48\textwidth, height=0.38\textwidth]{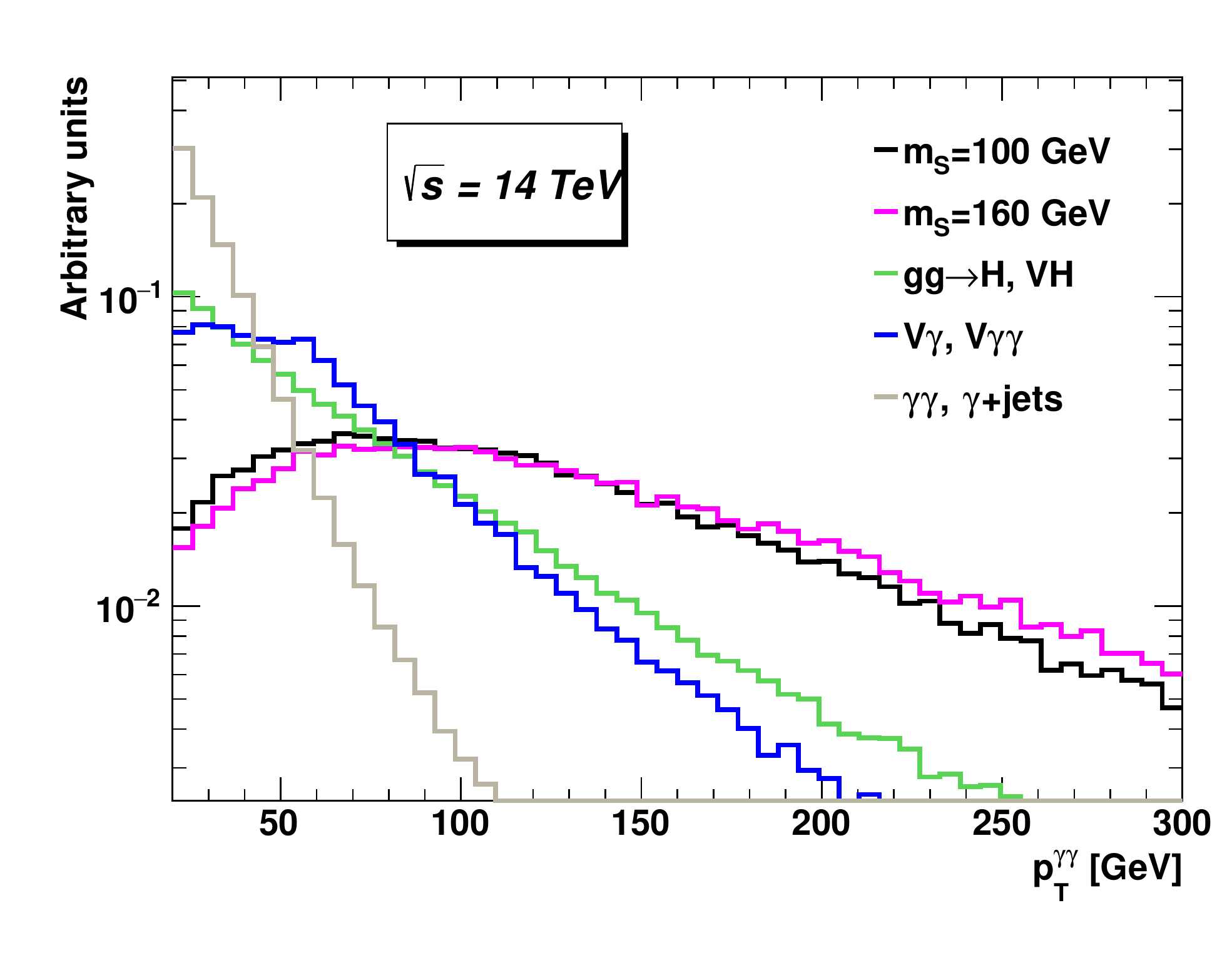}
\vfill
\includegraphics[width=0.48\textwidth, height=0.38\textwidth]{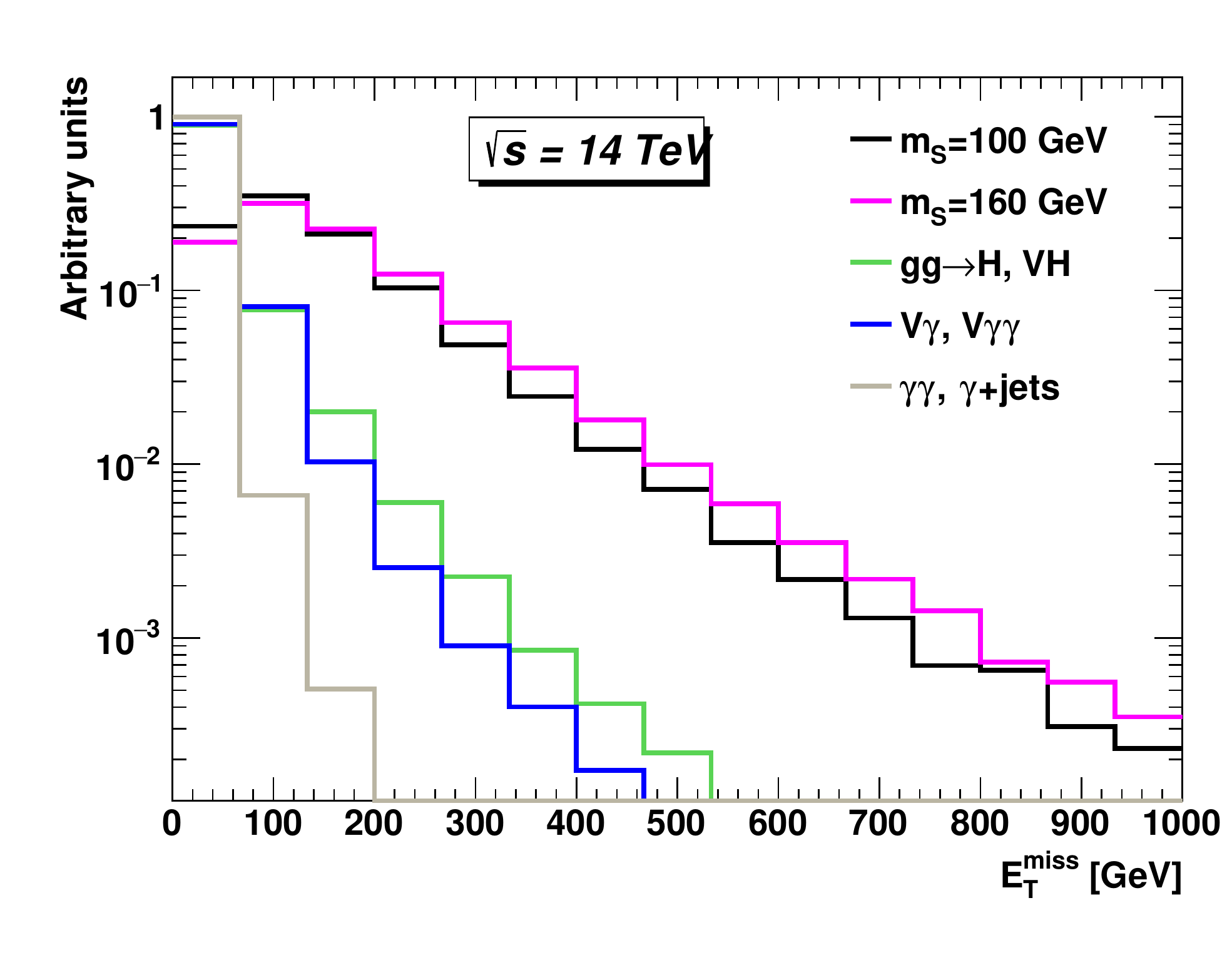}
\hfill
\includegraphics[width=0.48\textwidth, height=0.38\textwidth]{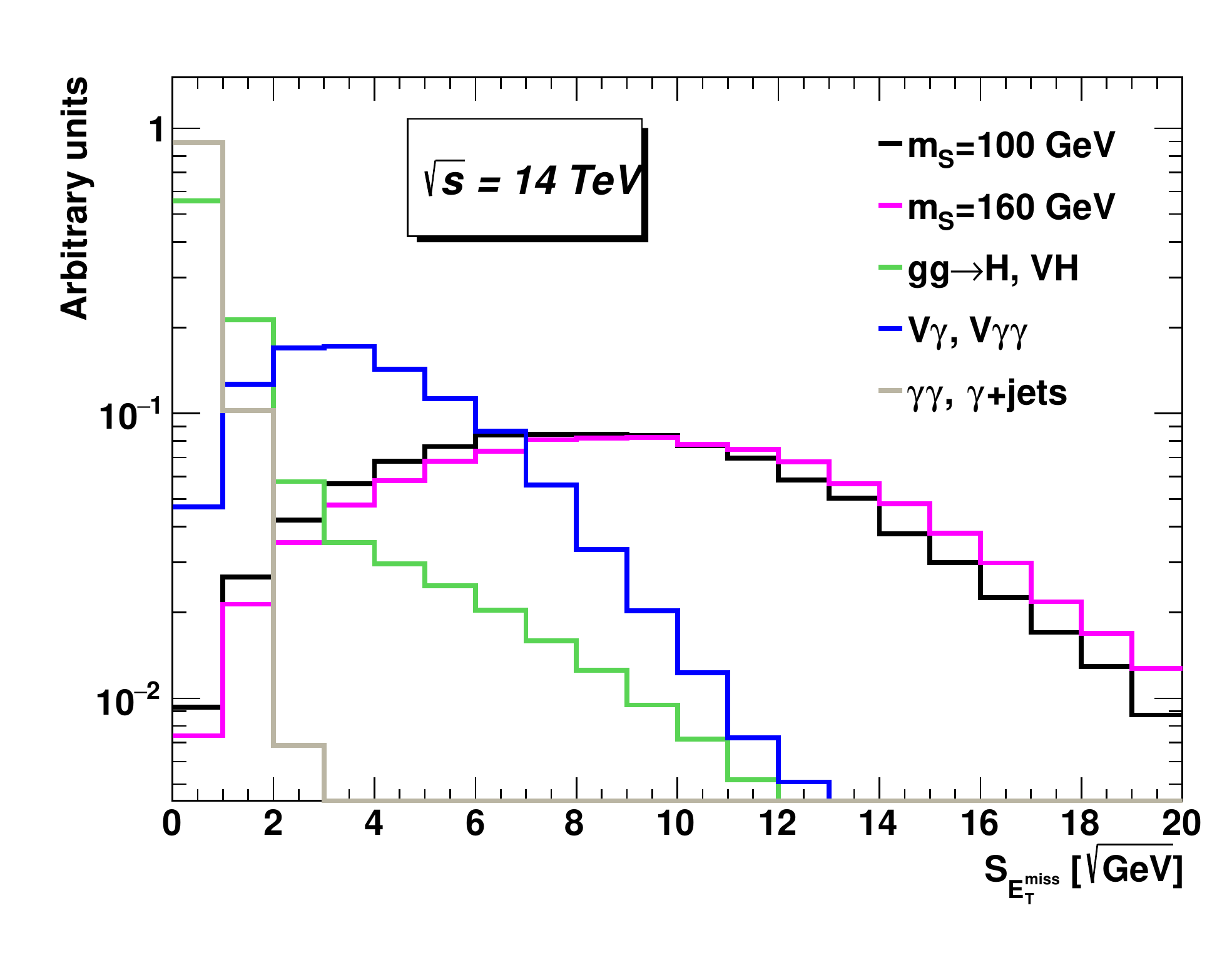}
\caption{Normalized distributions for the signal and the backgrounds at $\sqrt{s}=14$ TeV. Here, we show the transverse momentum of the leading photon $p_T^\gamma$ (top left), the transverse momentum of the diphoton system $p_T^{\gamma\gamma}$ (top right), missing transverse energy $E_T^\textrm{miss}$ (bottom left) and the $S_{E_T^\textrm{miss}}$ defined in eq.(\ref{SEtmiss}) (bottom right). The color coding is a follows; SM Higgs processes are shown in green, $V\gamma$ and $V\gamma\gamma$ are shown in blue, $\gamma\gamma$+jets are shown in gray. We show here the signal for $m_{H^0} = 100$ GeV (black) and $m_{H^0} = 160$ GeV (rose).}
\label{fig:distributions14TeV}
\end{figure}

\begin{figure}[h]
\includegraphics[width=0.48\textwidth, height=0.38\textwidth]{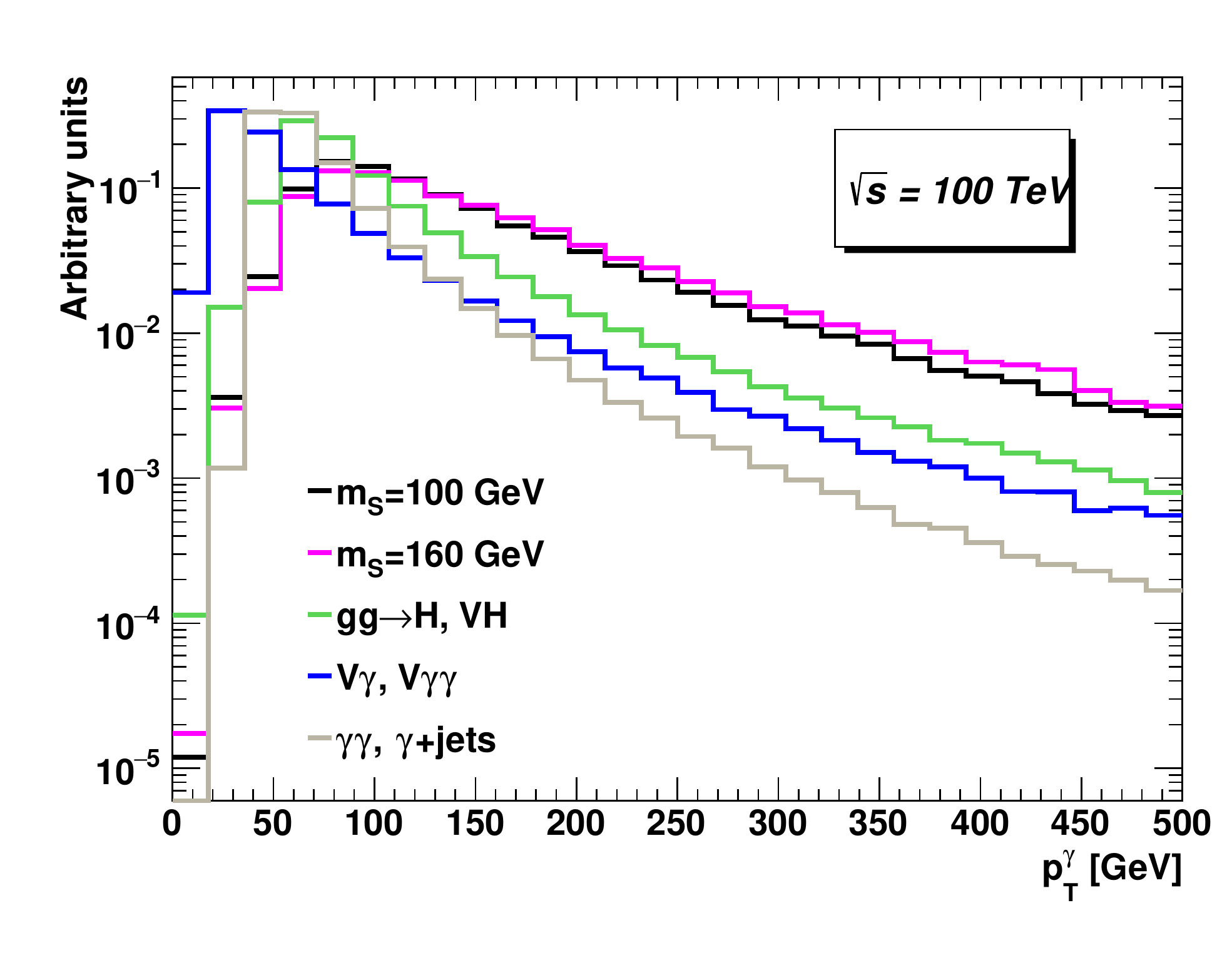}
\hfill
\includegraphics[width=0.48\textwidth, height=0.38\textwidth]{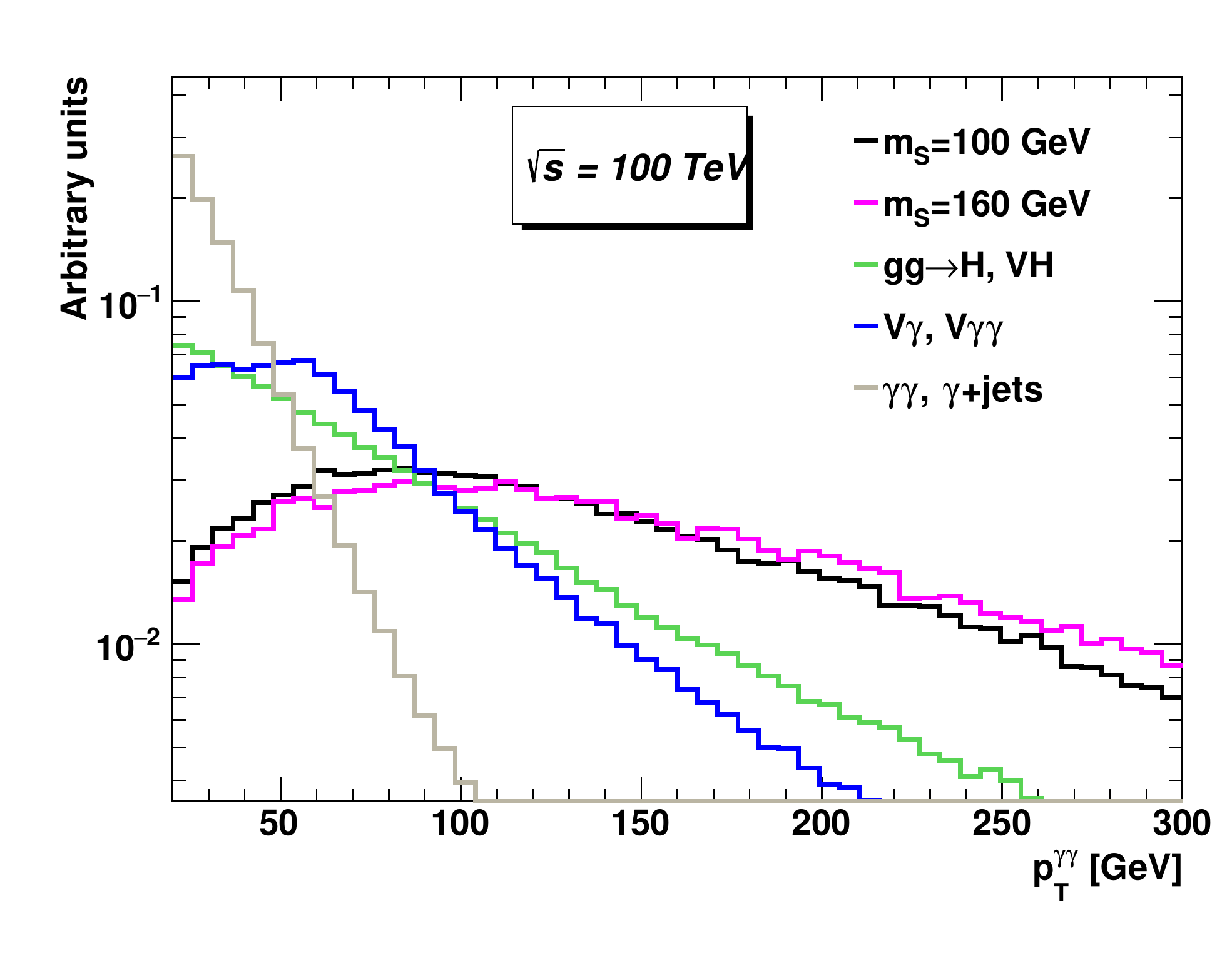}
\vfill
\includegraphics[width=0.48\textwidth, height=0.38\textwidth]{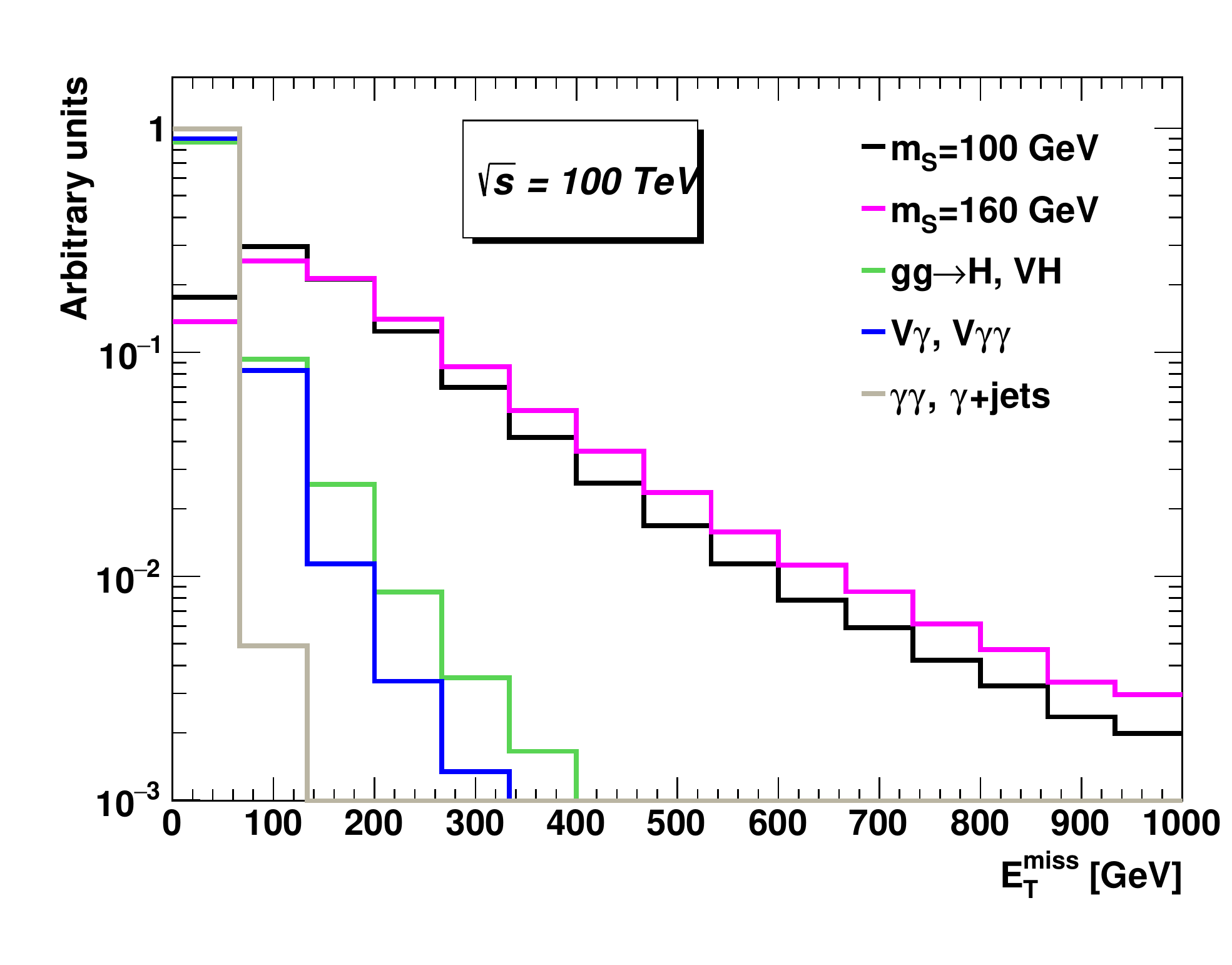}
\hfill
\includegraphics[width=0.48\textwidth, height=0.38\textwidth]{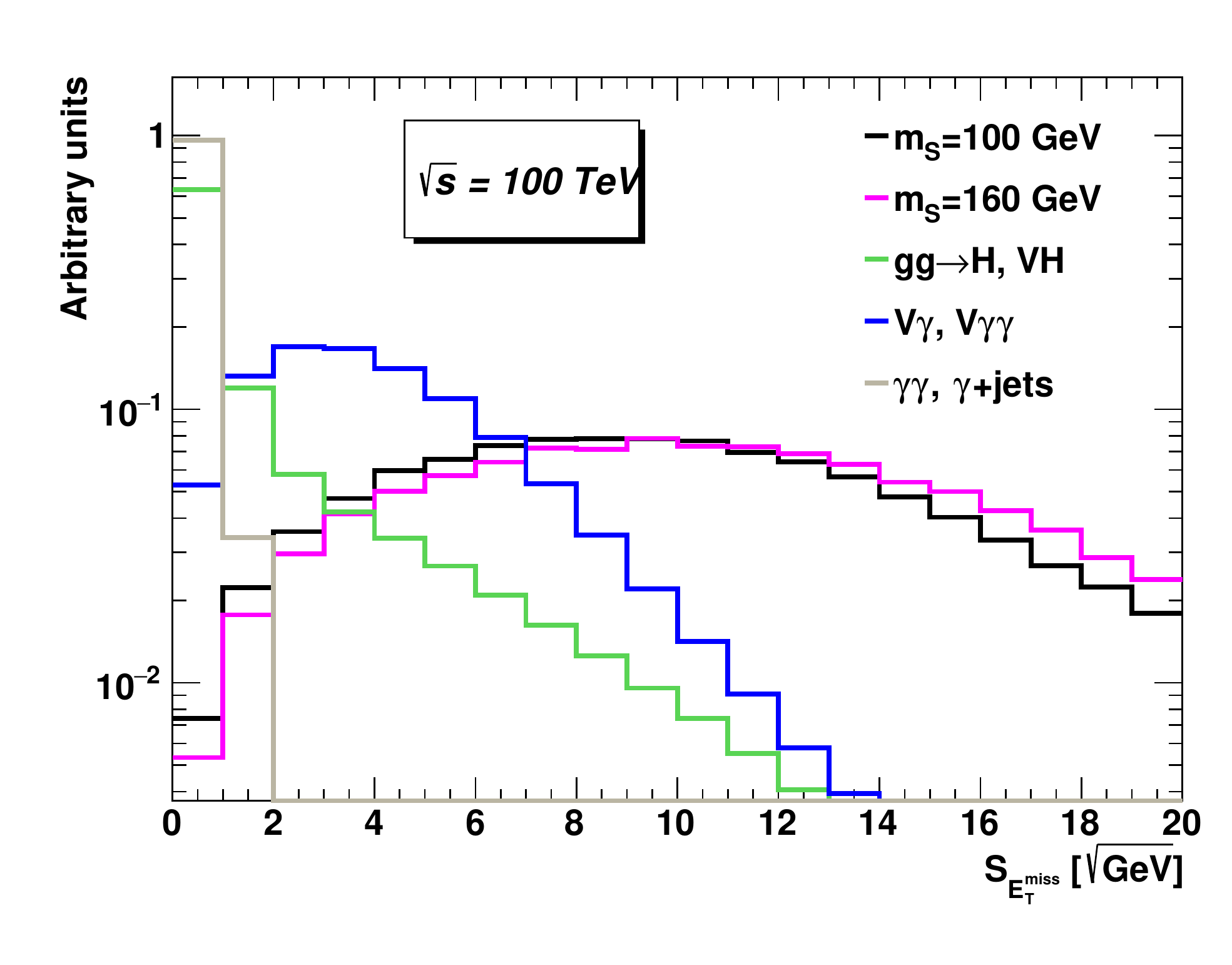}
\caption{Same as in Fig.~\ref{fig:distributions14TeV} but for $\sqrt{s}=100$ TeV.}
\label{fig:distributions100TeV}
\end{figure}

\section{Results and Discussion}

In Figs.~\ref{fig:distributions14TeV} and~\ref{fig:distributions100TeV}, we display the normalized distributions for some key observables used in
the signal-to-background optimization. We can see that the $p_T^\gamma$ of the leading photon is stronger for the signal than in the backgrounds
with a slightly high peak value for the signal case. The transverse momentum of the diphoton system (top right panel of Figs.~\ref{fig:distributions14TeV}
and~\ref{fig:distributions100TeV}) is a good discriminator. This is can be understood as follows; the \emph{Higgs candidate} (reconstructed from the
two photons) is produced in association with heavy particles (resulting in a hard missing transverse energy spectrum) and therefore the corresponding
recoil imply a harder $p_T$ than in the backgrounds (especially SM Higgs backgrounds and $\gamma\gamma$+jets). The same observation applies to the
$E_T^\textrm{miss}$ (bottom left panel). The $S_{E_T^\textrm{miss}}$ shows a very important discriminatory power between the signal and the backgrounds.
The condition used by the ATLAS collaboration to define the mono-Higgs signal region ($S_{E_T^\textrm{miss}} > 7$) can be considered as an optimum.
This is clear because requiring higher values for $S_{E_T^\textrm{miss}}^\textrm{min}$ will not only reduce the backgrounds but also diminish the signal.
We report on a difference between the results of our work and those in the ATLAS paper regarding the $S_{E_T^\textrm{miss}}$ and $p_T^{\gamma\gamma}$
variables; in the ATLAS paper, $\gamma\gamma$ and $\gamma$+jets events can still have some contribution to these variables (in the hard region) due to
the presence of pile-up events (which are not taken into account in our analysis). However, the number of events is still not very important; e.g.
about $10$ events for $S_{E_T^\textrm{miss}} > 7$ at $\sqrt{s}=13$ TeV and $\mathcal{L}=36.5~\textrm{fb}^{-1}$. We can assign the differences in the
modeling to an additional systematic uncertainty (see below).

\begin{figure}[h]
\includegraphics[width=0.48\textwidth, height=0.38\textwidth]{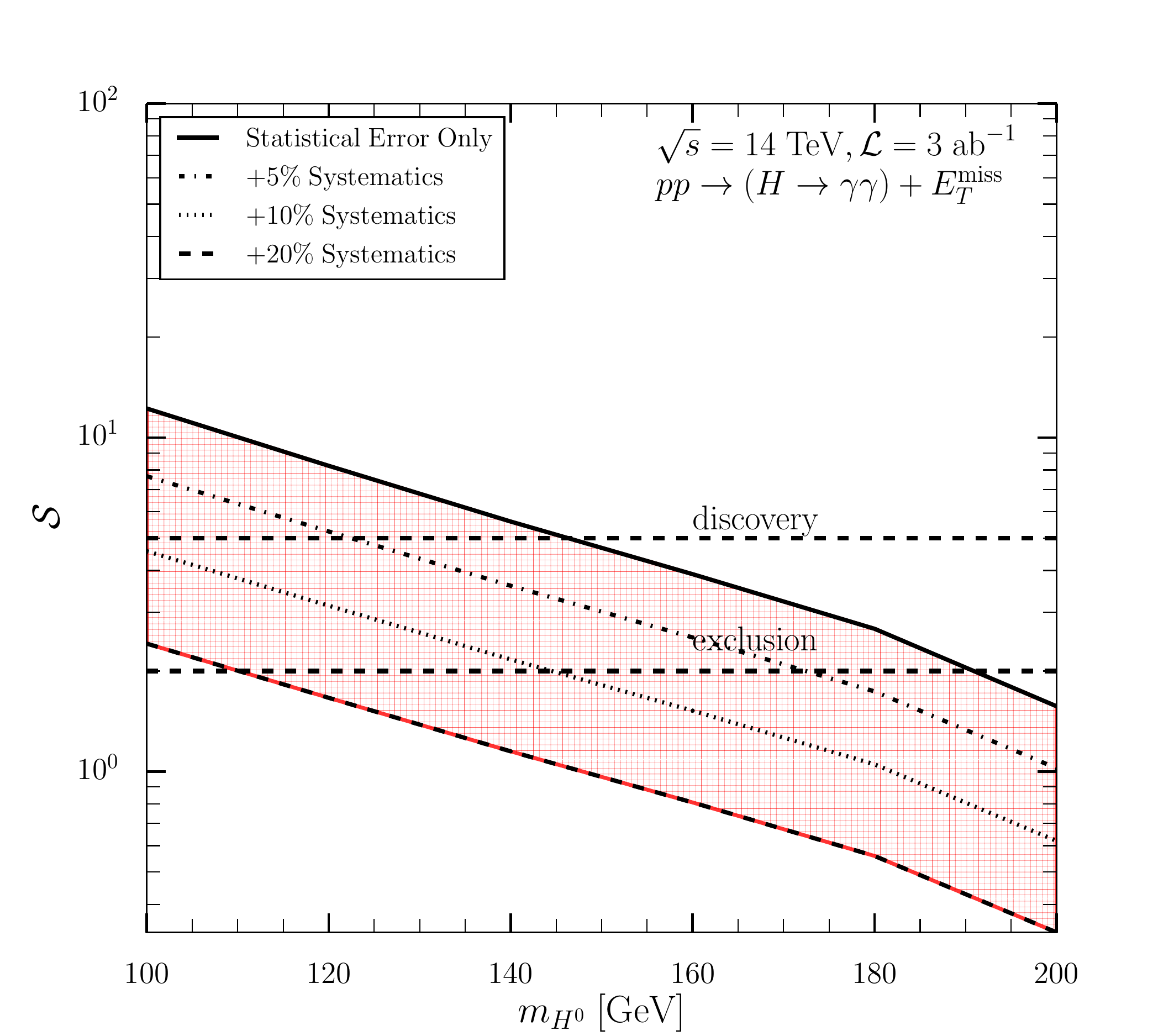}
\hfill
\includegraphics[width=0.48\textwidth, height=0.38\textwidth]{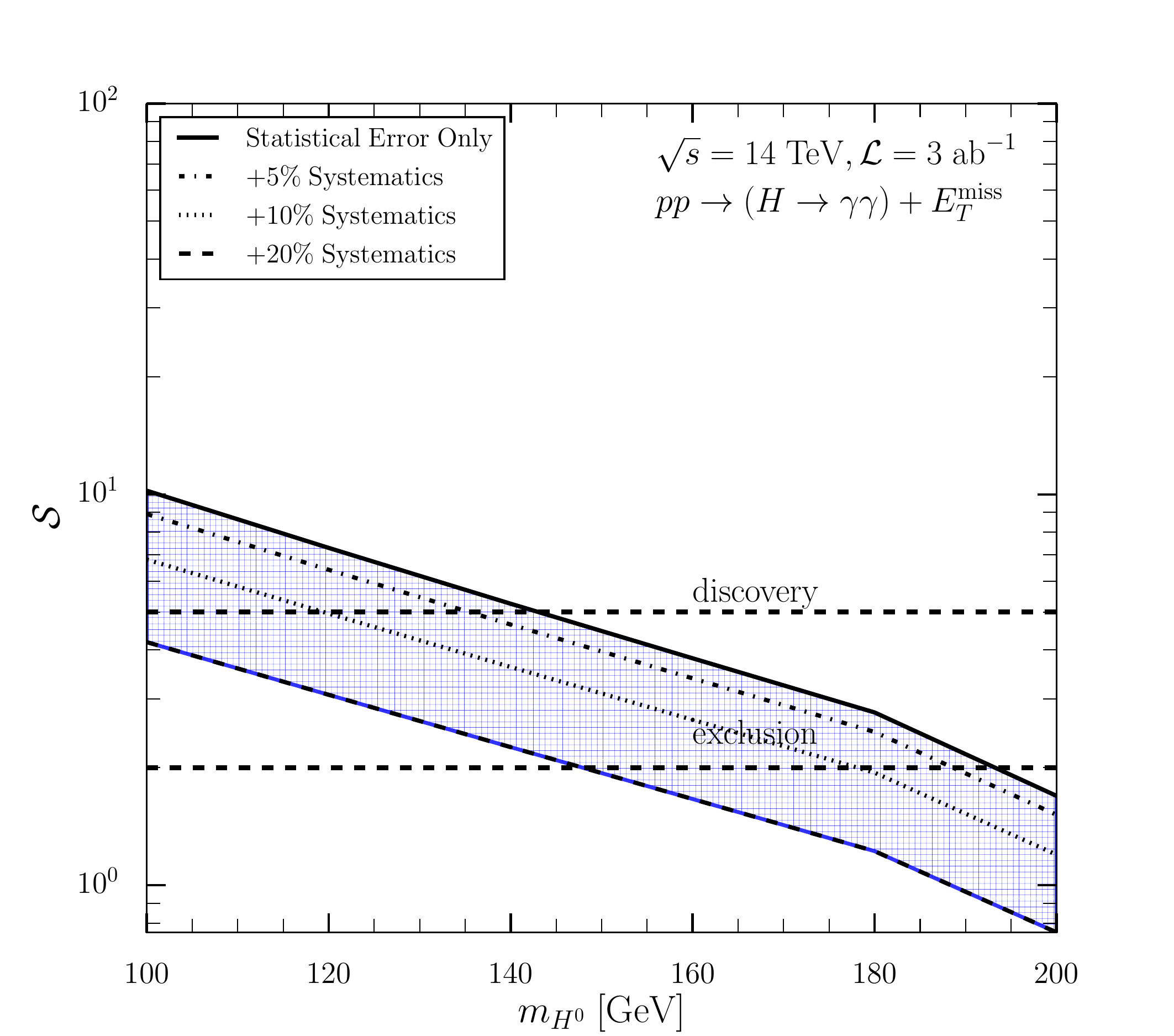}
\vfill
\includegraphics[width=0.48\textwidth, height=0.38\textwidth]{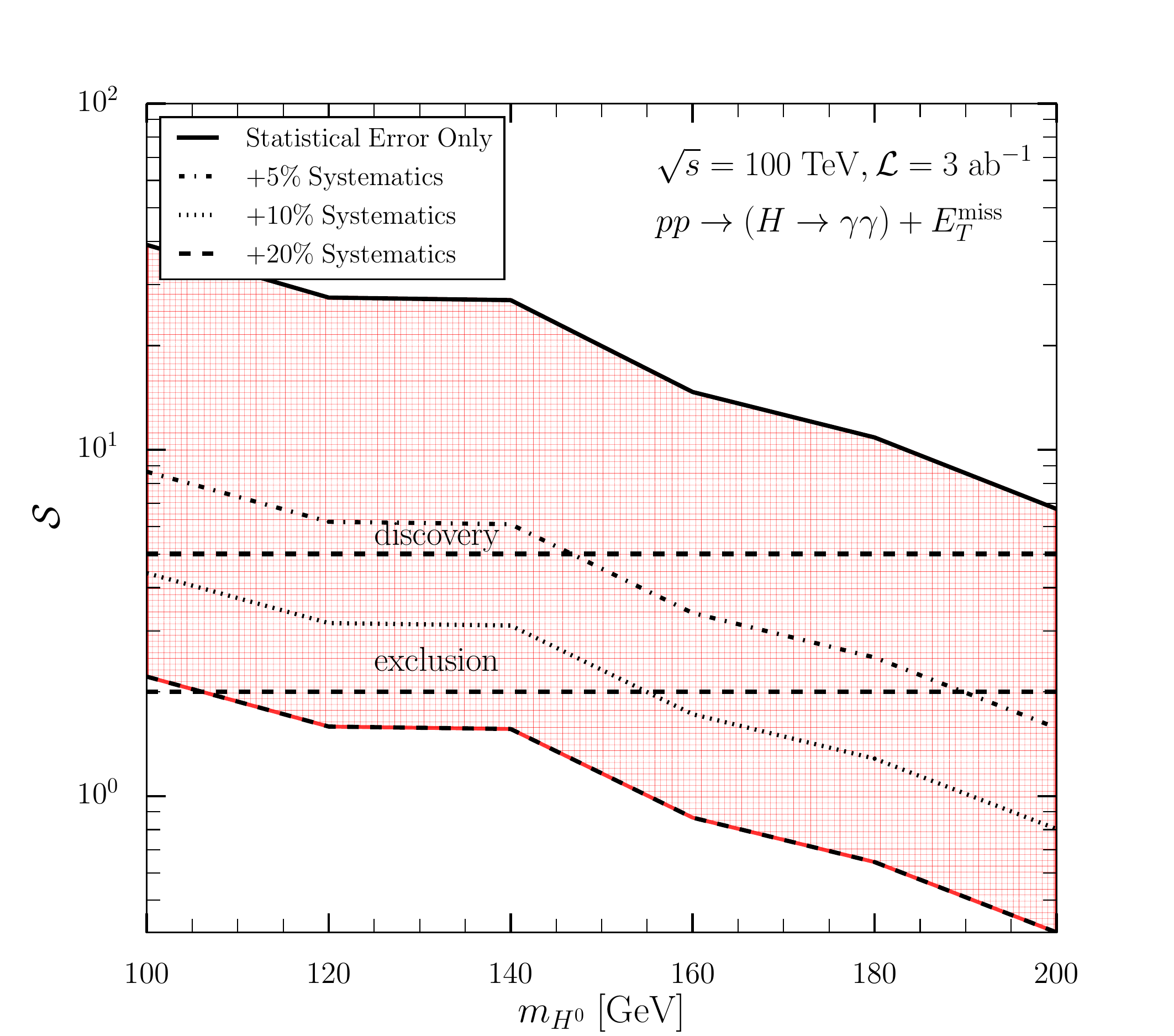}
\hfill
\includegraphics[width=0.48\textwidth, height=0.38\textwidth]{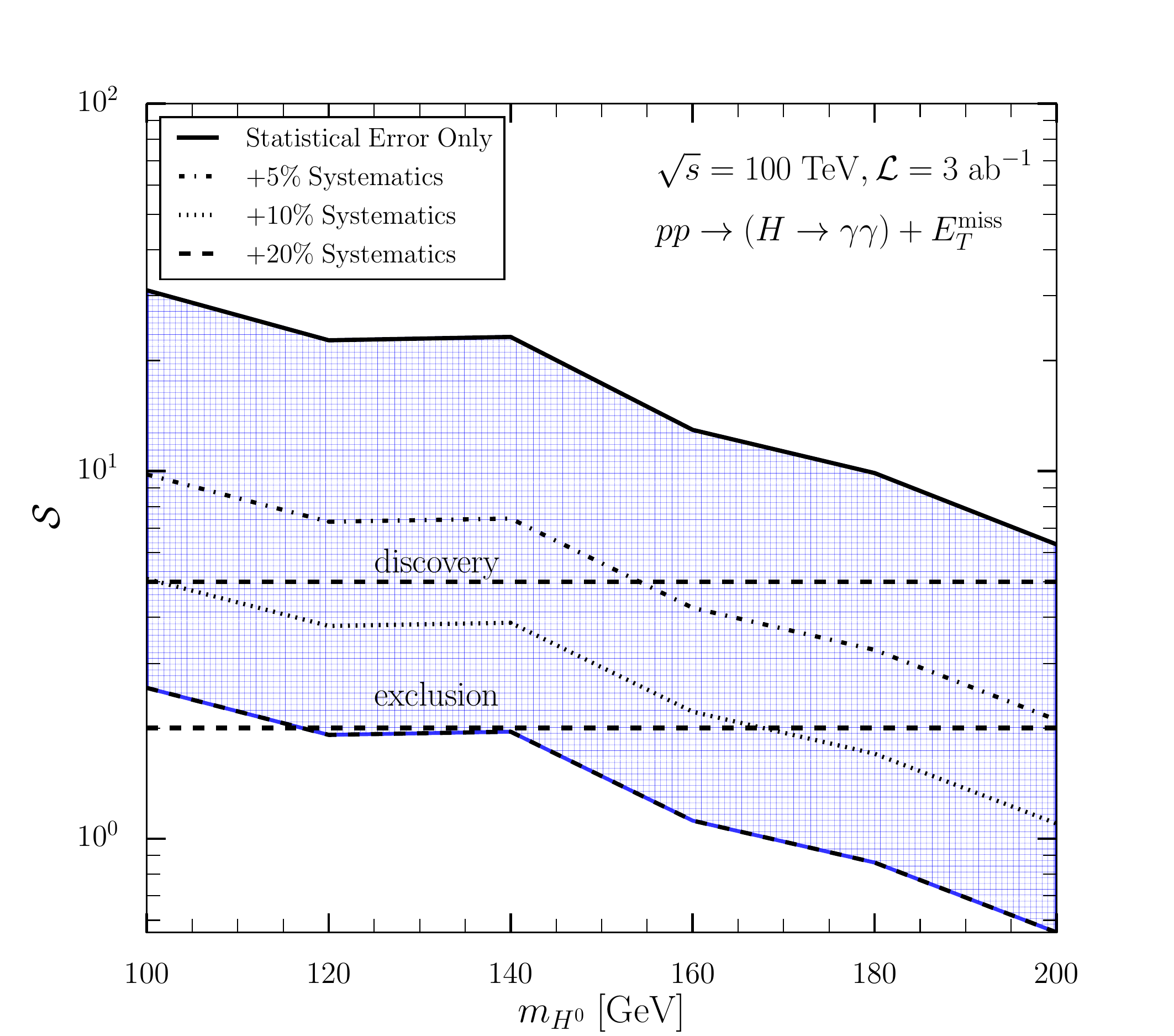}
\caption{Signal significance as a function of the dark Higgs mass $m_{H^0}$ at $\int \textrm{d}t \mathcal{L} = 3000~\textrm{fb}^{-1}$ in
the ATLAS mono-Higgs signal region (left panels) and after the tight selection (right panels) for the LHC at $\sqrt{s}=14$ TeV (top panels)
and for the FCC-hh at $\sqrt{s}=100$ TeV (bottom panels). In all the panels, the solid line correponds to the significance with the inclusion
of the statistical uncertainty only while dot-dashed, dotted and dashed lines include a $5\%$, $10\%$ and $20\%$ systematic uncertainty (summed
in quadrature with the statistical error in the corresponding signal region).}
\label{fig:significance-mass}
\end{figure}

\begin{figure}[h]
\begin{center}
\includegraphics[width=0.75\textwidth]{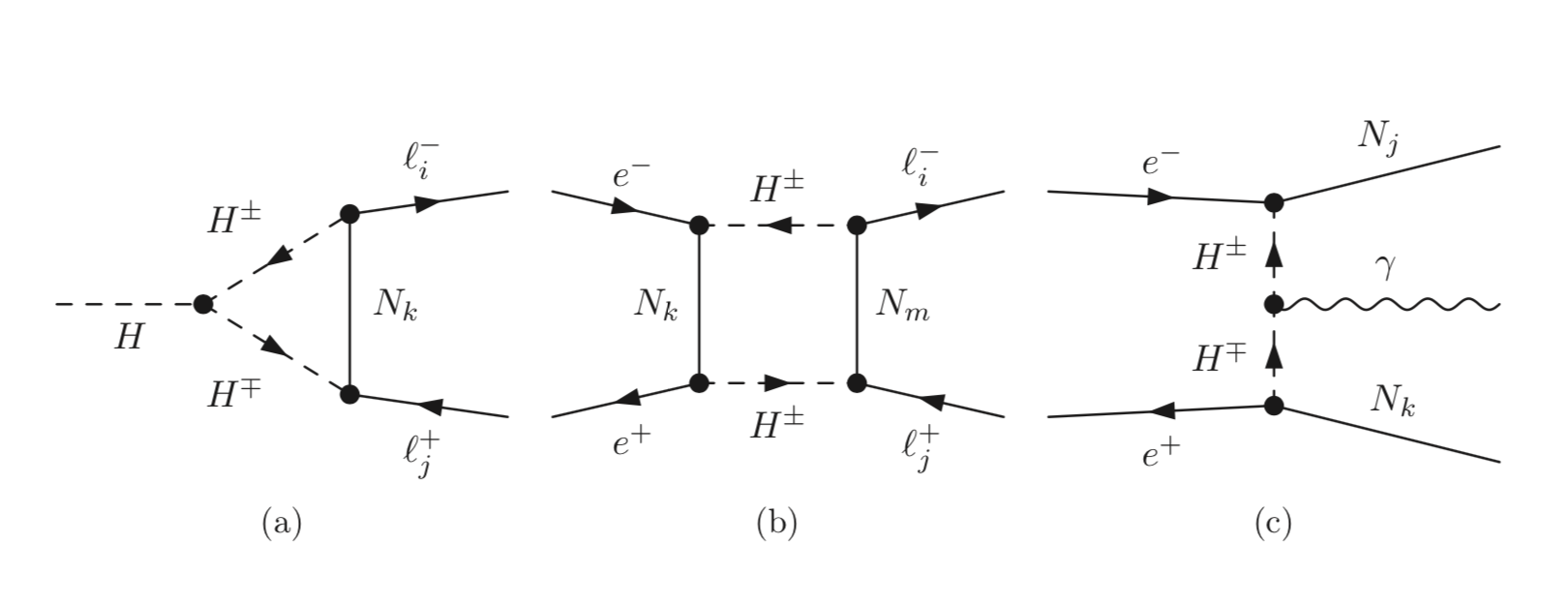}
\end{center}
\caption{Feynman diagrams of processes sensitive to the new Yukawa couplings. we show here the $H\to \ell_i \ell_j, i\neq j$ (diagram a),
the one-loop induced $e^+ e^- \to \ell^+_i \ell^-_j$ (diagram b) and an example of direct production of Majorana fermions (diagram c).}
\label{fig:observables}
\end{figure}

To quantify the discovery potential of the signal, we estimate the significance defined by~\cite{Cowan:2010js, Coleppa:2017lue}

\begin{eqnarray}
\mathcal{S}&=&\sqrt{2}\left[(s+b)\log\left(\frac{(s+b)(b+\delta_b^2)}{b^2+(s+b)\delta_b^2}\right) - \frac{b^2}{\delta_b^2} \log\left(1+\frac{\delta_b^2 s}{b(b+\delta_b^2)}\right)\right]^{1/2},
\label{S}
\end{eqnarray}
where $s$ and $b$ refer to the number of signal and background events respectively, and $\delta_b = x b$ is the uncertainty on the background events.
Before discussing the results of our sensitivity projections, we comment on the possible sources of systematic uncertainties and their impact
on background contribution. First, there are uncertainties related to missing higher order corrections and PDF+$\alpha_s$. Uncertainties due
to scale variations are usually determined by varying the renormalization and factorization by a factor of $2$ in two directions resulting in
an envelope composed of nine possible variations (assuming no correlations with the PDF uncertainties). These uncertainties on the SM Higgs backgrounds
are small due to the high precise calculations ($2.5$-$6\%$ for $m_{\gamma\gamma}/\textrm{GeV} \in [110:160]$). Following the recommendation
of PDF4LHC working group~\cite{Butterworth:2015oua}, PDF+$\alpha_s$ uncertainties can be estimated by combining both the variations of the same
PDF set (used in the calculation of the cross section) with the variations due to alternative PDFs\footnote{according to the recommendations of
PDF4LHC, the central PDF set is \textsc{Nnpdf30} while the two alternatives are \textsc{Ct10} and \textsc{Mmht}}. The size of the envelope spanned
by all the variations define the uncertainty due to PDF+$\alpha_s$. In the signal region, such uncertainties are very small for SM Higgs backgrounds
and can be of order $1.2$-$2.5\%$~\cite{Aaboud:2017uak}. An additional component of the theory uncertainty comes from the calculations of
$H\to\gamma\gamma$ branching ratio which is of order $1.73\%$~\cite{deFlorian:2016spz}. The uncertainties on the non-resonant backgrounds can be
larger than on the resonant backgrounds. In the analysis of~\cite{Aaboud:2017uak}, they were estimated directly from data and were of order $0.1$-$9.8\%$
in the $110 < m_{\gamma\gamma}/\textrm{GeV} < 160$ region. On the other hand, there are three major experimental uncertainties in the
$\gamma\gamma+E_T^\textrm{miss}$ final state; Luminosity, Photon identification efficiency and pileup reweighting. The total uncertainty on the
background contribution including both resonant and non-resonant processes was estimated by the ATLAS collaboration to be about
$15\%$\footnote{The ATLAS collaboration reported on the total error without specifying the contribution of systematic uncertainties. However,
the CMS collaboration~\cite{Sirunyan:2018fpy} reported both the contribution of the statistical error which is the dominant one and the systematic
error to the total uncertainty. In the signal region (defined as the High-$p_T^\textrm{miss}$ in the CMS paper), the total systematic uncertainty
is $\simeq 1.6\%$.}. In this work, we compute the signal significance taking into account statistical uncertainty only or statistical uncertainty
in addition to a systematic uncertainty of order $5\%$, $10\%$ and $20\%$. \\
In Fig.~\ref{fig:significance-mass}, we plot the significance of the signal process at $\mathcal{L} = 3000$ fb$^{-1}$ as a function of the dark Higgs mass $m_{H^0}$ for both the HL-LHC and FCC-hh. We show the significance in the ATLAS mono-Higgs signal region (left panels) and in the signal region defined in our paper by the tight selection (right panels). We can see that masses up to $140$ ($160$) GeV can be probed at the LHC (FCC-hh) if one assumes a $5\%$ of total error.

\section{Conclusions and Outlook}
In this work, we carried out a complete study of the mono-Higgs signature in the scotogenic model
in the limit of degenerate scalars with a focus on the $\gamma\gamma$ final state at both the LHC-HL and FCC-hh. After revisiting the
collider constraints from LEP and LHC run-II, we have shown that a considerable region of the parameter space
is still allowed which is already excluded in general scenarios. Using the most significant benchmark points, we have
shown that this model can be probed at the LHC-HL and the FCC-hh in the $H(\to\gamma\gamma)+E_T^\textrm{miss}$ channel
with $3$ ab$^{-1}$ of integrated luminosity. The final state we considered has a small rate compared to the other production
mechanisms of Majorana DM in the model, however, due to the cleanliness of the $\gamma\gamma$ decay channel and the high efficiency of photon
identification at hadron colliders, we have shown that it can be used to probe the model with the compressed spectrum. In summary, we have
found that scalar masses up to $150$ ($160$) GeV can be probed at the LHC (FCC-hh) assuming a $5\%$ systematic uncertainty. We stress out,
however, that these results can be significantly improved by the use of multivariate techniques such as boosted decision tree or neural
networks and, by including other decay channels of the SM Higgs boson with larger branching fractions. We point out that the importance
of the Mono-Higgs signature to probe the scalar coupling $\lambda_L$ which can't be probed using e.g. mono-jet searches of DM.

In the limit of compressed spectrum, i.e., for $\lambda_5 \simeq 0$; and for $m_{H^\pm} = 95$ GeV, the dark scalars decay exclusively
to a SM lepton (charged lepton or neutrino) and a Majorana fermion. Therefore, the mono-Higgs analysis itself is blind to the absolute values of
the new Yukawa couplings as well as to the number of Majorana fermions with mass below the scalar dark Higgs ($m_{N_k} < m_{H^0}$). This
conclusion can apply to all the production channels of dark scalars at hadron colliders. It is worth to investigate the potential of other
channels and observables to pin down such parameters. Below, we discuss briefly some methods to determine the new Yukawa couplings:

\begin{itemize}

\item \emph{Higgs flavor violating decays}. The SM Higgs boson is expected to undergo lepton flavor violating decays in the scotogenic
model. These decays are one-loop induced with the exchange of a charged scalar and Majorana fermions (Fig.~\ref{fig:observables}-a).
The ATLAS and the CMS collaborations have been searched for these decays channels at $\sqrt{s}=8\oplus 13$ TeV (see e.g.~\cite{Aad:2016blu,
Sirunyan:2017xzt}). The null results were used to put severe limits on the LFV Higgs decays, i.e $\textrm{BR}(H\to \mu\tau) < 0.25\%$
and $\textrm{BR}(H\to e\tau) < 0.61\%$~\cite{Sirunyan:2017xzt}. Possibly observing one or more of these decays channels can be used to
constrain one or several combinations of the new Yukawa coupling~(These processes are also quadratically dependent on
$g_{HH^\pm H^\pm} \propto \lambda_3$ which can still have large values).

\item \emph{Precision measurement of lepton pair production at Lepton Colliders}. In the scotogenic model, a pair of charged leptons
($e^+ e^- \to \ell^+_i \ell^-_j$) can be produced with decent rate at center-of-mass energies below or above the $Z$-boson pole (Fig.
~\ref{fig:observables}-b). Of these processes, the ones with $i \neq j$ are particularly interesting since they have almost zero cross
section in the SM. Therefore, measurement of both inclusive as well as differential rates in $\ell^+_i \ell^-_j$ can be used to extract
several combinations of the new Yukawa couplings.

\item \emph{Direct production at lepton colliders}. The production of inert scalars and Majorana fermions are the most sensitive channels
on the new Yukawa couplings (They can also be used as a model discriminators, see e.g.~\cite{Maturana-Avila:2019qph}). Production of
Majorana fermions (in association with photons, leptons or $Z$-bosons) either in the prompt mode or from the decays of dark scalars is
possible in the scotogenic model (in Fig.~\ref{fig:observables}-c we display an example of a Feynman diagram for $N_j N_k \gamma$ production).

\end{itemize}

\begin{acknowledgements}
A. Ahriche is supported by the Algerian Ministry of Higher Education and Scientific Research under
the PRFU Project No B00L02UN180120190003. A. Jueid would like to thank Robert V. Harlander for useful discussions, S. Banerjee for his help regarding the use of \textsc{Delphes} and F. Siegert for providing the correct setup of the calculation of the $\gamma$+jets cross section in
\textsc{Sherpa}. The work of A. Arhrib was supported by the Moroccan Ministry of Higher Education and Scientific Research
MESRSFC and CNRST: "Projet dans les domaines prioritaires de la recherche scientifique et du d\'eveloppement technologique": PPR/2015/6.
The work of AJ is sponsored by CEPC theory program and by the National Natural Science Foundation of China
under the Grants No. 11875189 and No.11835005. A. de la Puente would like to thank
the ICTP for support where part of this work has been done. The Feynman diagrams are draw using \textsc{JaxoDraw}~\cite{Binosi:2003yf, Binosi:2008ig}
and \textsc{FeynArts}~\cite{Hahn:2000kx}.
\end{acknowledgements}

\appendix

\section{Recasting of LHC searches of new physics at $\sqrt{s}=13$ TeV}
\label{sec:appendix}

In section~\ref{sec:LHC}, we studied the impact of the LHC searches of
new physics beyond the SM on the parameter space of the compressed scotogenic model. We discuss here the phenomenological setup of the event generation and a brief
description of the analyses used in the reinterpretation effort. These analyses, implemented in \textsc{Checkmate}, are listed in Table~~\ref{tab:constraints}.
\begin{itemize}

\item atlas\_conf\_2016\_050~\cite{ATLAS:2016ljb}: the ATLAS collaboration has been searched for new phenomena in the final state consisting of $1~\ell + (b) \textrm{jets} +E_T^\textrm{miss}$ at $13.3$ fb$^{-1}$ of luminosity. These searches were focused on the supersymmetric partner of the top quark, and also on DM production in association with a pair of top quarks. Upper bounds on the stop quark mass (for different assumptions regarding its decay branching ratios) were put. Furthermore, limits on DM simplified models were obtained and presented on a plane of DM mass and pseudo-scalar mediator mass for a coupling $g_{DM} = 3.5$.

\item atlas\_conf\_2016\_066~\cite{ATLAS:2016fks}: Using a dataset corresponding to $13.3$ fb$^{-1}$ of luminosity, searches of new physics in the final state consisting of one photon, jets and large $E_T^\textrm{miss}$ is performed by the ATLAS collaboration. These searches were used to probe supersymmetric models with gauge-mediated supersymmetry breaking, where neutralinos decay into a photon and a gravitino. Limits were put on the mass of a degenerate gluino state; i.e. $m_{\tilde{g}} > 1800$ GeV for a large range of neutralino (the Next-to-Lightest Supersymmetric Particle -NLSP- which is a mixture of higgsino and bino) masses and $m_{\tilde{g}} > 2000$ GeV for high neutralino mass.

\item atlas\_conf\_2016\_076~\cite{ATLAS:2016xcm}: A search of stop pair production and DM production in association with $t\bar{t}$ has been performed using $13.3$ fb$^{-1}$ of integrated luminosity. This search targeted final states composed of $2$ charged leptons, jets and large $E_T^\textrm{miss}$. From the non-observation of a beyond the SM signal, $95\%$ CL model-independent upper limits on the visible cross section were obtained (they vary between $0.38~\textrm{fb}$ and $1.18~\textrm{fb}$ depending on the analysis strategy).

\item atlas\_conf\_2017\_060~\cite{ATLAS:2017dnw}: Using a larger dataset corresponding to $36.1$ fb$^{-1}$, a search for new physics in the mono-jet final state ($1~\textrm{jet} +E_T^\textrm{miss}$) is performed. Good agreement with the SM expectation was observed. As a consequence, exclusion limits on different models (with pair-produced weakly interacting DM candidates, large extra dimensions, and SUSY particles in several compressed scenarios) were obtained.

\item atlas\_1704\_03848~\cite{Aaboud:2017dor}: A search for new physics in the mono-photon final state ($1~\gamma +E_T^\textrm{miss}$) with dataset corresponding to $\mathcal{L} = 36.1$ fb$^{-1}$ was performed. $95\%$ CL limits were put on models with $s$-channel pseudo-scalar mediators, effective field theory models and on the production of a heavy $Z'$ decaying into $Z(\to \nu\nu)+\gamma$.

\item atlas\_1709\_04183~\cite{Aaboud:2017ayj}: A search of a stop pair production was performed using the final state $0~\ell + (n\ge4)~\textrm{jets} +E_T^\textrm{miss}$ at luminosity 36.1 $fb^{-1}$. The null searches were used to put exclusions limits on the top-squark and neutralino masses.

\item atlas\_1712\_02332~\cite{Aaboud:2017ayj}: The final state consisting of ($2$-$6$) \textrm{jets} $+E_T^\textrm{miss}$ at the luminosity $36.1$ fb$^{-1}$ recorded by the ATLAS detector, was used to search for squarks and gluinos. $95\%$ CL lower limits on gluino masses ($m_{\tilde{g}} > 2.03$ TeV) and squark masses ($m_{\tilde{q}} > 1.55$ TeV) were placed.

\item atlas\_1712\_08119~\cite{Aaboud:2017leg}: the final states with two low-momentum leptons and missing transverse momentum is used to search for electroweak
production of SUSY particles in scenarios with compressed mass spectra at the luminosity $36.1$ fb$^{-1}$ recorded by ATLAS. Exclusion limits on SUSY particles
masses are established.

\item atlas\_1802\_03158~\cite{Aaboud:2018doq}: Using $36.1$ fb$^{-1}$ of luminosity, photonic signatures (single photon and diphoton) in association with large $E_T^\textrm{miss}$ are considered to look for SUSY particles production in generalized models of gauge-mediated supersymmetry breaking. using 36.1 $fb^{-1}$ recorded by ATLAS. In these models, lower limits of $2.15$ TeV, $1.82$ TeV and $1.06$ GeV are set on the masses of gluinos, squarks and a degenerate set of winos, respectively (for any value of the bino mass less than the mass of these produced states).

\item cms\_sus\_16\_025~\cite{CMS:2016zvj}: the final state of two low-momentum opposite-sign leptons and missing transverse momentum in events recorded by
CMS at luminosity $12.9$ fb$^{-1}$ of data collected at $13$ TeV, to search for many new physics model candidates. The observed data yields are compatible with the SM predictions,
and upper bounds of $175$ GeV on charginos and the next-to-lightest neutralino are set, with a mass difference of $7.5$ GeV with respect to the lightest neutralino.

\item cms\_sus\_16\_039~\cite{Sirunyan:2017lae}: Using the data recorded by CMS at $13$ TeV and luminosity $35.9$ fb$^{-1}$, the final state of multileptons is considered to search for
neutralinos and charginos that are weakly produced. In simplified SUSY models, these negative searches were interpreted as exclusions on the mass
interval $180$-$1150$ GeV.

\item cms\_sus\_16\_048~\cite{Sirunyan:2018iwl}: Using the same CMS dataset, the final state consisting of two low-momentum, oppositely charged leptons with missing
transverse momentum is used to search for new physics. Negative searches results implied exclusions on the wino-like masses up to $230$ GeV
for $20$ GeV mass difference relative to the lightest neutralino, and the higgsino-like masses are excluded up to 168 GeV for the same mass difference. In addition,
the top squark masses up to $450$ GeV are excluded for a mass difference of $40$ GeV relative to the lightest neutralino.

\end{itemize}

Several processes in the scotogenic model are sensitive to these searches. These processes lead to different final states; $\ell^+\ell^- + E_T^\textrm{miss}, 1 \gamma+E_T^\textrm{miss}$, mono-jet, and $1\ell + \textrm{jets} + E_T^\textrm{miss}$ among others. First, Charged Higgs boson pair production will lead to a final state composed primarily of 2 isolated charged leptons and a large $E_T^\textrm{miss}$. In some cases, where one charged lepton escapes the detection, this final state can be triggered as a $1\ell + \textrm{jets} + E_T^\textrm{miss}$ where the jets are produced in initial state radiation. For small mass splittings ($\Delta_{H^\pm} = m_{H^\pm} - m_{N_k})$, the missing transverse energy triggered by the Majorana fermion is even larger and thus gives high sensitivity. Production of a CP-odd (CP-even) dark scalar in association with a charged Higgs boson ($pp\to H^0 H^\pm$) leads exclusively to $1\ell + \textrm{jets} + E_T^\textrm{miss}$. We also considered the mono-$V$ process with $V=W, Z$ which contributes to a final state composed of multi-jets ($n \geq 2$) and large transverse missing energy. On the other hand, mono-photon and mono-jet processes are also possible in this model. For mono-jet production, we generated $S^0 S^0 + n$ jets ($S^0=H^0, A^0$) using \textsc{Madgraph5\_aMC@NLO}~\cite{Alwall:2014hca} with jet multiplicity up to $3$ jets. We matched these inclusive samples using the \textsc{MLM} matching scheme~\cite{Mangano:2006rw}. \textsc{Pythia} 8.155~\cite{Sjostrand:2007gs} was used for showering and hadronization. We have added by hand the PDG codes of the three Majorana fermions (which should be considered as invisible particles) to the HCAL modules of the \textsc{Delphes} card.

\bibliographystyle{unsrt}


\end{document}